# Il rapporto tra ICT e PMI italiane e le problematiche economico-organizzative dell'OS.

Laura Anna Ripamonti
*D.I.Co. Dipartimento di Informatica e Comunicazione - Università degli Studi di Milano*

## SOMMARIO

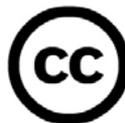

## 1. Introduzione

Scopo del progetto OS4E è comprendere se e come le tecnologie *open source* possano essere proficuamente adottate come base per il core business di un'impresa che si occupa di *system integration* orientata al mercato delle imprese italiane di dimensioni medio-piccole appartenenti prevalentemente – ma non esclusivamente - al settore petrolchimico e meccanico (sostanzialmente l'indotto delle grandi raffinerie).

Trattandosi di un approccio decisamente innovativo (ad es. la redditività nel tempo dei modelli di business basati sull'*open source* continua ad essere un dibattito aperto sotto molti punti di vista), è difficile fornire una risposta ragionevolmente sensata al quesito proposto basandosi esclusivamente sull'indagine di esperienze analoghe.

Non potendo, perciò, prescindere da un apporto analitico da parte della ricerca, è indispensabile contenere la complessità del problema, analizzandolo nelle sue principali componenti. Ponendoci in un'ottica aziendale, possiamo percepire alcune criticità principali, che possono essere così riassunte:

1. comprendere quale sia il modello di adozione delle PMI per quanto riguarda le tecnologie OS. Questo implica poter correttamente affrontare la domanda potenziale;
2. comprendere come confrontare e selezionare componenti F/OSS da includere nelle propria offerta (si tratta in particolare di soluzioni da integrare in un software infrastrutturale);
3. comprendere le reciproche influenze tra valutazione delle qualità funzionali e non dei prodotti OS e modello di adozione;
4. comprendere quali difficoltà economiche e organizzative sia necessario affrontare per gestire adeguatamente la transizione verso un modello che affonda le sue radici culturali nell'idea di "*gift for reputation*" piuttosto che in quella di "*good for money*".

L'analisi di queste problematiche implica l'adozione di un approccio multidisciplinare, in grado di coniugare sinergicamente competenze tecniche ed economico-organizzative.  Tale analisi, a sua volta, deve essere radicata in un'appropriata analisi dei risultati di ricerca fino ad ora pubblicati relativamente a queste problematiche, in modo ipotizzare e costruire modelli comportamentali e decisionali ragionevolmente "allo stato dell'arte".

Il presente *report* si è occupato di fornire una base analitica e concettuale su cui costruire poi ipotesi di risposta per i precedenti punti 1 e 4. In particolare il lavoro è suddiviso in due macro-sezioni, la prima delle quali si focalizza sull'analisi del complesso e conflittuale rapporto che lega la PMI italiana all'uso strategico delle ICT (capp. 2, 3 e 4), riportando i dati maggiormente significativi che è stato possibile rinvenire in letteratura o in *survey* pubblicate (in molti casi i dati riportati fanno riferimento a uno specifico testo o a una specifica indagine che si sono rivelati particolarmente rappresentativi della problematica generale e/o della posizione di un certo numero di autori). La la seconda sezione invece riporta alcune considerazioni riguardanti sia la sostenibilità economica dei modelli di business basati sul paradigma OS, che le problematiche (e le opportunità) legate alla gestione di community "ibride" di sviluppatori OS.

# PARTE 1

# IL RAPPORTO TRA LE PICCOLE E MEDIE IMPRESE E LE TECNOLOGIE DI COMUNICAZIONE E INFORMAZIONE (ICT)

## 2. Le piccole e medie imprese (PMI) nell'economia moderna.

> *Non è il più forte che riesce a sopravvivere, né il più intelligente,*
> *ma chi riesce maggiormente ad adattarsi al cambiamento.*
> (Charles Darwin)

L'unica certezza sull'ambiente competitivo nel quale le imprese si troveranno immerse nell'immediato futuro è che il grado di incertezza nel quale si troveranno ad operare sarà sensibilmente maggiore rispetto a quello odierno (cfr. Fig.1).

Sembra un'osservazione scontata, ma implica conseguenze più profonde di quanto non appaia in prima battuta. La crescita di incertezza, infatti, richiede a un'impresa, nel maneggiare le innovazioni tecnologiche, di confrontarsi con qualcosa che è in continua evoluzione e cambiamento. Non solo: questa continua evoluzione va a insistere su di un contesto nel quale molteplici differenti *trend* stanno (o sono già) convergendo per creare condizioni che stanno inesorabilmente sovvertendo le tradizionali regole del gioco competitivo.

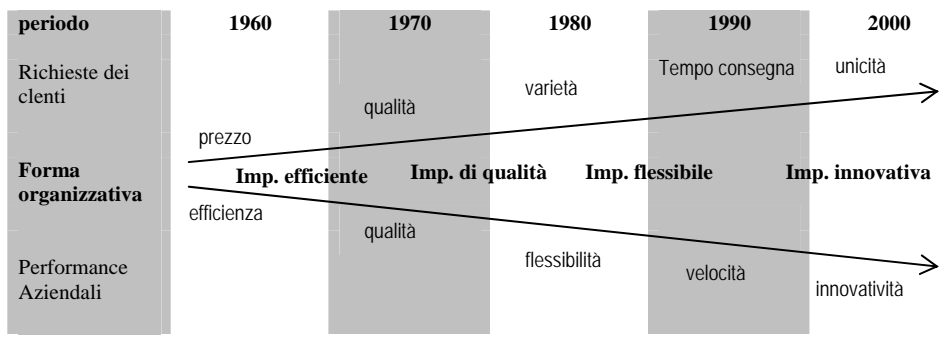

**Figura 1 -** Relazione tra l'evoluzione della domanda di mercato e le competenze richiesta alle imprese per sostenere la competizione (Boer, 2002).

Le ICT (Tecnologie di Informazione e Comunicazione) sono state – di fatto – tra i principali fattori di innesco del momento di discontinuità che la moderna economia sta attraversando, e stanno, in misura sempre maggiore, pervadendo ogni momento della nostra vita quotidiana e del nostro modo di fare business. L'attuale momento congiunturale, inoltre, è caratterizzato da bassi tassi di inflazione, caduta degli investimenti, generale senso di "stanchezza": tutte queste condizioni congiuntamente contribuiscono all'integrazione, attraverso il commercio e gli investimenti, di economie trans-nazionali.

In questo panorama la velocità di risposta, la flessibilità e l'adattabilità divengono vieppiù i fattori critici per la sopravvivenza: tutti questi aspetti, tuttavia, non fanno altro che enfatizzare, tra le varie forme organizzative "vincenti", il paradigma "rete" supportata – ovviamente - dalle ICT.

Infatti, è facile verificare come, a fronte della perdurante fase di variabilità dei mercati internazionali, stia emergendo una sorta di doppia convergenza tra forme imprenditoriali: le piccole e medie imprese (PMI) si organizzano vieppiù in reti, sviluppando forme di servizi condivisi (spesso mutuati dai modelli della grande impresa), mentre le grandi organizzazioni cercano di ricreare con le consociate e i sub-fornitori relazioni di collaborazione tipiche di forme quali i distretti industriali.

Ma come si pone la PMI (Piccola e Media Impresa) italiana in tale contesto?

**2.1. PMI, network e altre classificazioni esotiche**

L'economia italiana dipende fortemente dalle (reti di) PMI (cfr. Tab.1, i cui dati rispecchiano la situazione fino al 2003, che non è sensibilmente differente dall'attuale). La PMI, infatti, impiega circa l'80% della popolazione attiva italiana, con le seguenti proporzioni: le piccole imprese (meno di 100 dipendenti) impiegano il 70% dell'intera forza lavoro (nel resto dell'OCSE questa percentuale si aggira attorno al 35-40%), mentre le medie imprese ne occupano un ulteriore 10% (contro il 20% nel resto dell'OCSE). È quindi evidente l'importanza della buona salute delle PMI (che è tra l'altro cruciale per l'innovazione e per la diffusione dell'immagine e della presenza italiana all'estero) per il sistema economico italiano.

Le teorie economiche sono a volte miopi, e le previsioni basate su di esse rischiano ogni tanto di rivelarsi errate, anche se difficilmente tali errori vengono prontamente riconosciuti. La distribuzione dimensionale delle imprese nei Paesi occidentali è uno dei più popolari tra questi "errori": secondo le previsioni, le imprese di minori dimensioni sarebbero state soppiantate da quelle più grandi nella maggior parte dei settori industriali. La produzione di nicchia a basso costo e il *subcontracting* per la grande impresa erano viste come le uniche chance di sopravvivenza per la PMI. Le grandi imprese si sarebbero imposte come *leader*, e il loro modello – basato su economie di scala e di scopo, sul potere finanziario e sul controllo del mercato - sarebbe diventato il modello vincente per qualsiasi tipo di business (Galbraith, 1963).

**Tabella 1 -** Imprese in Italia – Fonte: Sirmi per Assintel, 2002, Assinform, 2003

| **Imprese in relazione al territorio** | | | | |
|---|---|---|---|---|
| | **Micro imprese[1]** | **PMI[2]** | **Medio – grandi[3]** | *Totale* |
| Nord Ovest | 973.852 | 63.068 | 1.595 | *1.038.515* |
| Nord Est | 698.228 | 51.815 | 913 | *750.956* |
| Centro | 702.424 | 34.686 | 586 | *737.696* |
| Sud | 659.560 | 21.816 | 299 | *681.675* |
| Isole | 304.356 | 8.107 | 111 | *312.574* |
| **Totale Italia** | ***3.338.420*** | ***179.492*** | ***3.504*** | ***3.521.416*** |
| **% di imprese in relazione al territorio** | | | | |
| | **Micro imprese** | **PMI** | **Medio – grandi** | *% totale* |
| Nord Ovest | 93.8 | 6.1 | 0.15 | *29.5* |
| Nord Est | 93.0 | 6.9 | 0.12 | *21.3* |
| Centro | 95.2 | 4.7 | 0.08 | *20.9* |
| Sud | 96.8 | 3.2 | 0.04 | *19.4* |
| Isole | 97.4 | 2.6 | 0.04 | *8.9* |
| **Totale Italia** | ***94.8*** | ***5.1*** | ***0.10*** | ***100*** |

[1] da 1 a 9 dipendenti, [2] da 10 a 199 dipendenti, [3] più di 200 dipendenti

Sebbene la teoria economica non abbia mai specificato quanto tempo avrebbe impiegato – dopo tutto la PMI esisteva da sempre - questo modello di convergenza per concretizzarsi, il risultato finale era ben chiaro, persino in termini statistici: alcune grandi imprese si sarebbero conquistate una posizione centrale nell'encomia, e sarebbero state circondate da un pugno di medie e piccole aziende totalmente impossibilitate ad agire autonomamente.

Questa idea della convergenza delle grandi imprese e della loro conseguente "vittoria sul resto del mondo" era basta su due assunzioni principali: la prima aveva a che fare con gli aspetti legati *all'efficienza* (costi e prezzi bassi erano considerati la discriminante sulla base della quale si decidevano le sorti di vincitori e perdenti nel gioco competitivo, a prescindere dalle differenze intercorrenti tra i prodotti, le strutture dei mercati, le tecnologie impiegate e i bisogni dei clienti). Tuttavia il semplice contenimento di costi e prezzi era una strategia abbastanza facilmente perseguibile fino a quando la conoscenza e la tecnologia si evolvevano a una molto velocità contenuta (o trascurabile): le grandi imprese erano così in grado di imporre il ritmo dei cambiamenti, assicurandosi così ottime opportunità di crescita - la replicazione del medesimo processo produttivo e dei medesimi prodotti era un prerequisito per poter beneficiare delle tanto osannate economie di scala. Le PMI, effettivamente, non avevano possibilità: se "*anybody was free to buy any car, but a black model T*", Ford era l'unico possibile vincitore.  Le statistiche dimostrarono che questa assunzione era effettivamente vera per molti tra i settori industriali maggiormente popolari (nel giro di poche decadi, ad esempio, il numero dei produttori di automobili negli Stati Uniti passò da più di 200 a meno di una dozzina). La seconda assunzione aveva a che fare con l'informazione, la conoscenza e il potere. Le argomentazioni a sostegno di questa assunzione sono più sottili. Le dimensioni e il potere di mercato non erano una semplice conseguenza delle economie di scala, ma anche una condizione che contribuiva a rafforzare il vantaggio competitivo delle grandi realtà organizzative. L'integrazione verticale consentiva ai grandi *player* di controlla la catena del valore, e – soprattutto – le risorse e le competenze chiave erano sviluppate internamente.  Conseguentemente le relazioni con partner e fornitori esterni esistevano solo sulla base di considerazioni tattiche e opportunistiche.  Le grandi imprese avevano il controllo sulla conoscenza "cruciale", e potevano perciò permettersi di decidere la velocità con la quale introdurre nuove tecnologie, nuovi prodotti e nuovi processi (e quindi trarne beneficio), assumendosi rischi estremamente bassi. Malgrado oggi possa sembrare *naïve*, questa visione ha dominato la letteratura economica per decenni ed è definitivamente crollata solo relativamente di recente.

Anziché essere rimpiazzate dalla grande impresa, le PMI sono cresciute di numero in molti Paesi occidentali; anziché giocare un ruolo marginale, le PMI hanno condotto il gioco dell'innovazione e della competizione sia in settori maturi che innovativi; anziché dipendere dalle grandi imprese, le PMI hanno aperto la via dell'innovazione tecnologica.

A questo punto, un eterogeneo gruppo di economisti, professionisti e sociologi avanzarono delle convincenti spiegazioni per il successo delle PMI: la flessibilità e la specializzazione possono essere più efficaci della dimensione, l'innovazione continua di prodotto e processo possono essere più importanti del puro costo nel discriminare tra perdenti e vincenti, le risorse tangibili e quelle intangibili si integrano in maniera maggiormente appropriata se focalizzate e supportate da un'opportuna motivazione nel lungo periodo.

Nella ricerca di spiegazioni convincenti per la persistenza e la crescita delle PMI gli economisti e i sociologi congiuntamente sottolinearono che la maniera di competere delle PMI è "differente".  Anziché adottare un atteggiamento "tutti contro tutti", le PMI cooperano per mettere a fattor comune i rispettivi punti di forza.  Questa cooperatività non è un caso: spesso si tratta di atteggiamenti basati sulla *fiducia reciproca* e rinforzati da meccanismo di "*quasi-mercato*" radicati da lungo tempo nella storia di specifici territori.

È interessante notare che la manifestazione di queste reti di collaborazione è stata oggetto di attento studio, tant'è vero che in letteratura esiste un numero pressoché infinito di possibili tassonomie per classificare le reti di imprese[1].

---

[1] Una tra le più note classificazioni è attribuibile a Rosenfeld, anche se essa si limita a tracciare una distinzione abbastanza grossolana tra reti formali "hard", nelle quali le imprese cooperano sul mercato o nello sviluppo/produzione di nuovi prodotti, e reti "soft" informali, nelle quali le imprese collaborano esclusivamente per risolvere problemi comuni o per scambiare informazioni e skill specifici. Thorelli (1986) e Ibarra (1992), invece, si sono concentrati soprattutto sugli aspetti economici, classificando le reti come forme organizzative intermedie tra il mercato e la gerarchia, caratterizzate da interdipendenze a livello comunicativo e da scambi orizzontali di risorse.  Powell osserva che le reti di imprese sono intrinsecamente dotate di una flessibilità molto maggiore delle gerarchie, grazie a una maggiore diffusione delle informazioni rilevanti e a un minor numero di vincoli. Piuttosto rilevante è anche il lavoro di Forsgren et al. (1995), che evidenzia il legame tra la forte interattività e collaborazione tra le imprese appartenenti a reti e la correlazione positiva intercorrente tra l'intensità di interazione e l'aumento di fiducia reciproca. Huggings (1997, 1998) evidenzia tre differenti tipi di reti: *networks of information*, of *knowledge*, e of *innovation*, poste in ordine crescente di interattività e difficoltà implementativa.  Hanssen-Bauer & Snow (1996) studiarono la funzione cognitiva intrinseca alle reti, che deriva dal mix di interessi, fiducia e linguaggio comune che li caratterizza.  Naturalmente non possiamo dimenticare il contributo fondamentale di  Priore &

## 3. PMI e ICT: ostacoli all'adozione.

Come abbiamo visto, le PMI creano relazioni interorganizzative perché realizzano immediatamente i benefici – immediati e futuri – che ne possono derivare. Le reti di PMI, perciò, emergono spontaneamente, sulla base di una serie di decisioni che non necessariamente dipendono da come la relazione si evolverà nel tempo (ad esempio il concetto di distretto industriale è archetipico di questo modello di sviluppo: piccole imprese altamente specializzate, localizzate in un'area geografica generalmente ben definita, che cooperano molto strettamente e sono sostenute dalle istituzioni pubbliche locali fino a diventare leader mondiali nel loro settore).

A prima vista, dunque, i servizi e le infrastrutture di rete offerte dalle ICT dovrebbero essere ideali per le (reti di) PMI: (apparentemente) accessibili a costi piuttosto contenuti, flessibili e adeguate per lo scambio informativo. E, in effetti, un uso saggio delle ICT si è dimostrato efficace per aumentare la competitività, e offre alle PMI - grazie ad Internet - l'opportunità di competere "ad armi pari" con organizzazioni di dimensioni ben maggiori. In particolare il World Wide Web e l'e-mail (che già nel 1997 Spectrum definiva rispettivamente "*key service of the Information society*" e "*leading environment for interactive environment*") offrono alle PMI l'opportunità - a costo praticamente nullo – di condividere informazioni senza barriere di tempo nè geografiche. Tuttavia l'evidenza empirica suggerisce che l'adozione e l'uso di Internet da parte delle PMI progredisce lentamente.

**3.1 La posizione della ricerca scientifica sul rapporto tra PMI e ICT.**

Analizzando la letteratura scientifica che riguarda le imprese di dimensioni medio piccole[2] si possono fare alcune interessanti considerazioni, che riguardano sia la consistenza del dibattito scientifico intorno a questa tematica, che la *vision* strategica delle PMI nei confronti delle ICT, che le principali barriere all'adozione di ICT nella PMI, che le conseguenze della resistenza all'adozione delle ICT nella PMI.

---

Sabel (1984), che individua la doppia convergenza tra distretto industriale da una parte, e grande azienda decentralizzata dall'altra. In un lavoro abbastanza recente, Soda (1998) ha cercato di sistematizzare la teoria delle reti di imprese, individuando tre tipi fondamentali di reti in un continuum di assetti organizzativi ibridi tra l'impresa integrata a il mercato perfetto: reti *burocratiche*, *proprietarie* e *sociali*. Anche se meritevole di nota, il lavoro di Soda ha la pecca di non considerare alcune variabili potenzialmente rilevanti per una classificazione efficace delle reti di impresa, come la localizzazione geografica, il sistema di *inpu-output* e la struttura di *governance*. A questo proposito Bennet & Storper (1992) suggeriscono l'adozione di una matrice che considera anche queste ulteriori variabili per definire uno strumento operativo atto a classificare le reti.

[2] Si ricorda che la nozione di PMI così come viene intesa nel nostro Paese può differire anche considerevolmente – soprattutto per quanto concerne il fattore dimensionale – da quella adottata in contesti ad esempio anglosassoni.

Per quanto concerne il dibattito scientifico, si rileva che, malgrado le PMI siano una componente rilevante dell'economia globale (e non meramente italiana), la conoscenza riguardante i meccanismi che in esse regolano l'adozione delle ICT è abbastanza lacunosa, dato che il dibattito si è prevalentemente focalizzato sulle problematiche dell'allineamento IT nella grande impresa. Conseguentemente gli indicatori che misurano l'esistenza e la consistenza di una correlazione tra l'allineamento tecnologico e l'eventuale incremento di vantaggio strategico nella PMI non abbondano affatto.

In generale, comunque, da numerosi studi emerge che la PMI non dispone delle *risorse* (soprattutto a livello di competenze) necessarie a utilizzare in maniera veramente strategica le ICT, e che è spesso incapace di produrre strategie efficaci per regolare gli *investimenti* in ICT.

In letteratura si sottolinea anche la necessità di evitare fuorvianti generalizzazioni riguardo la PMI, dato che le *finalità e la metodologia di adozione* nelle imprese di ridottissime dimensioni e in quelle di dimensioni medio-piccole possono presentare casistiche e modelli di comportamento anche molto differenti.

Riguardo la problematica dell'adozione di ICT nelle PMI, la ricerca scientifica ci dice che quattro sono i principali fattori di ostacolo (specialmente per quanto riguarda l'interazione e la collaborazione inter-organizzativa):

1. <u>scarsa comprensione delle opportunità connesse alle tecnologie adeguate alle necessità delle PMI</u>: questo impedisce alle PMI di superare i gap di performance o di esplorare nuove opportunità;
2. <u>scarsa comprensione di come implementare adeguatamente le tecnologie</u>: questo comporta una scarsa capacità di perseguire efficienza, efficacia e innovazione;
3. <u>carenza di appropriati *skill* nelle risorse umane</u>: tipicamente una PMI dispone di uno staff ridottissimo (o non ne dispone affatto!) dedicato alle ICT;
4. <u>costo della tecnologia</u>

Inoltre, le PMI sembrano spesso mancare della *volontà e/o dalla capacità di dedicare tempo e risorse per colmare le lacune* sopra descritte.

Per valutare la potenziale gravità di questa resistenza all'adozione è importante capire quale sia l'impatto delle ICT sulle *performance* delle PMI. La notevole quantità di studi effettuati a questo proposito nella *grande impresa* riportano sia correlazioni positive che nulle: ciò indica che l'impatto delle ICT sulle *performance* non è diretto, ma mediato da altri fattori. Questo implica l'importanza, sottolineata in letteratura, di sviluppare un adeguato *fit* tra le strategie di sviluppo delle IT e quelle di sviluppo del business. Analoghe analisi non sono state effettuate in maniera altrettanto approfondita e sistematica nella PMI, anche se è possibile reperire alcuni interessanti spunti di riflessione. Ad esempio, alcuni ricercatori hanno rilevato che in genere *le imprese maggiormente innovative sono anche quelle che integrano maggiormente le ICT al loro business*, e che *le imprese che fanno un uso sofisticato delle ICT sono tendenzialmente quelle che hanno anche sviluppato una lungimirante strategia per la gestione tecnologica*. Altri ricercatori si sono concentrati sulle problematiche di allineamento tra strategia di business e strategia IT, proponendo anche approcci ad hoc orientati alla PMI. Tutti, comunque, concordano sulla criticità – sul piano delle performance – *dell'allineamento tra strategia e IT*.

Non solo, alcuni studiosi affermano che le *capacità manageriali*, la *conoscenza* e *l'esperienza* accumulate dal singolo piccolo imprenditore sono ciò che può "fare la differenza" nella capacità di sfruttare vantaggiosamente le ICT. È quindi di cruciale importanza fornire a questi imprenditori (o manager) gli strumenti necessari a sviluppare una profonda comprensione delle potenzialità insite nella tecnologie, affinché comprendano come esse possano assisterli nel lavoro quotidiano e nel business, e sviluppino coerenti piani di investimento.

Per finire, è interessante ricordare come nell'ambito scientifico sia ormai riconosciuta la necessità di "trattare" le tecnologie di comunicazione *Internet-based* (e quindi le ICT) come cosa a sé rispetto alle altre applicazioni basate sull'IT. Esse, infatti, operano una trasformazione sul business che è radicalmente differente da quella tradizionalmente attribuita all'IT, in quanto Internet è primariamente un *enabler* delle attività inter-organizzative e mette a disposizione un'infrastruttura informativa per molti versi paragonabile a quella telefonica (quando non addirittura strettamente integrata ad essa, come nel caso del Voice over IP – VoIP).

Riassumendo le considerazioni fin qui presentate, possiamo osservare che:

− un uso strategico delle ICT può avere un impatto positivo sia sulle performance che sulla competitività delle PMI;
− le IT in generale influiscono sul sistema interno di applicativi, mentre le tecnologie legate a Internet tendono ad avere un impatto positivo soprattutto sulle attività inter-organizzative;
− per poter trarre vantaggi reali e duraturi dall'uso delle ICT (impatti positivi sulle performance aziendali e sulla competitività) è necessario sviluppare un adeguato *fit* tra le strategie tecnologiche e di business;
− la profonda comprensione delle potenzialità insite nelle ICT da parte del management delle PMI è essenziale per lo sviluppo delle strategie;
− le PMI in genere non sono in grado di sviluppare questo tipo di vision strategica, a causa di mancanza di skill, conoscenza della tecnologia e costo della tecnologia stessa;
− il management beneficerebbe moltissimo di un'appropriata assistenza all'apprendimento delle potenzialità delle ICT.

Vediamo ora come si pone la PMI italiana rispetto all'adozione delle ICT.

**3.2 Stato dell'adozione delle ICT nella PMI italiana.**

Dalla classifica europea della penetrazione di Internet e delle ICT diffusa dal Ministero per l'Innovazione, risulta che l'Italia occupa uno degli ultimi posti, insieme alla Grecia e la Spagna, e che due italiani su tre possono essere considerati completi analfabeti nel campo delle ICT.

Questa situazione incide negativamente sull'innovatività e sulla produttività industriale del Paese. Negli ultimi anni L'Italia ha speso molto poco per le ICT (cfr. Fig. 2), malgrado circa il 7% delle imprese sia connesso ad Internet, ma questo non è una novità, infatti, a quanto risulta dai dati ufficiali, tra il 1992 e il 2001 l'Italia ha costantemente occupato le ultimissime posizioni nella classifica dei Paesi europei, sia per la spesa in tecnologie informatiche che per quella in tecnologie di telecomunicazione: rispettivamente il 2,02% e il 2,24% del PIL, totalizzando così una spesa pari al 4,26% del PIL, cioè l'antepenultimo Paese della UE.

Conseguentemente, ad esempio, in Italia la produttività è cresciuta del 1,67% per lavoratore, e del 2,28% all'ora, contro le medie europee del 2,1% e 3,59% rispettivamente.  Per finire, l'indagine ufficiale sottolinea che le imprese di telecomunicazioni italiane sono di dimensioni piuttosto piccole, il che contribuisce a indebolire ulteriormente il Paese nel campo delle ICT.

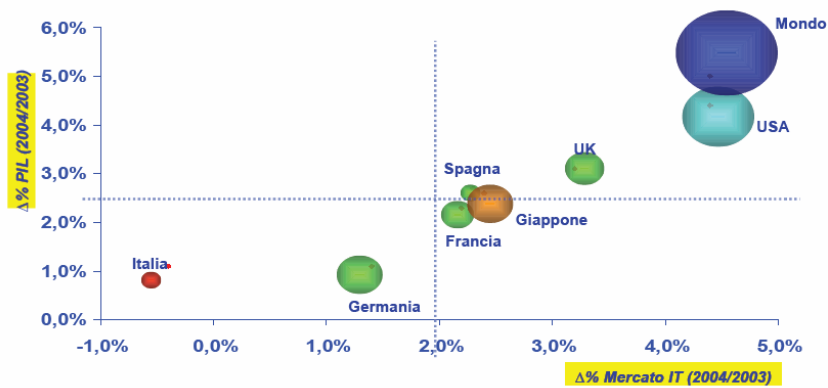

**Figura 2** - Fonte: Assinform/NetConsulting su dati OECD, FMI

I dati su base nazionale del Mistero sono confermati a livello locale dalle molte analisi condotte da Università, Camere di Commercio, Associazioni Industriali, ecc., dalle quali è interessante tentare di estrapolare le informazioni relative alle PMI.

In prima battuta, paragonando i dati di Fig.3 (relativi al numero di utenti Internet che si collega da casa e dal lavoro e al numero di potenziali utenti) e di Fig. 4 (che riporta la crescita tra il 1999 e il 2002 della penetrazione di Internet nelle PMI) con quelli riassunti in Tab. 1 ("Imprese in Italia"), risulta evidente che la maggioranza delle aziende possiede una connessione Internet, confermando il dato che indica il 72% delle imprese connesse a Internet.

Data l'entità considerevole del numero di imprese connesse ad Internet, è interessante osservare il peso della PMI sul mercato IT (cfr. Fig.5): i dati sono maggiormente rilevanti se considerati congiuntamente a quelli di Fig.6, Tab.2 e Tab.3 e successivamente comparati con quelli indicati dal Ministero. L'analisi condotta dal Ministero, infatti, riporta un dato globale per la spesa ICT pari al 4,26% del PIL tra il 1999 e il 2001, situazione dettagliata attraverso Fig.5, che evidenzia come solo una piccola percentuale (lo 0,89% del PIL nella migliore delle ipotesi) di questa spesa sia attribuibile alla PMI, che è per contro la base dell'economia italiana e impiega circa l'80% dell'intera forza lavoro del nostro Paese.

Inoltre Tab.2 evidenzia l'esistenza di un "gap tecnologico" tra PMI localizzate in territori differenti (tradizionalmente il Sud è sempre stato prono a un certo svantaggio economico, e questa situazione viene confermata dal livello di spesa in ICT per lavoratore – Tab.3) e una progressiva contrazione di tale spesa nel corso degli anni 2000, 2001 e 2002 (questo dato è confermato anche da Fig. 7, che pone in evidenza la differente dimensione dell'investimento all'interno dei differenti gruppi dimensionali di imprese).

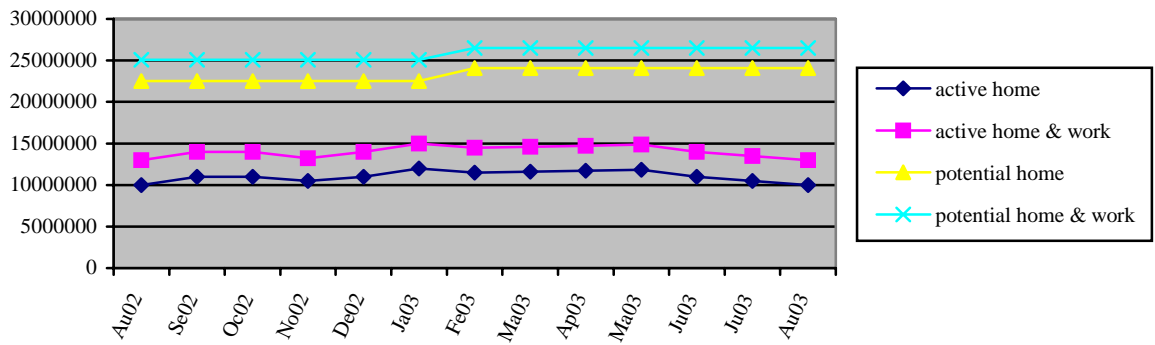

**Figura 3 -** Internet in Italia – Fonte dei dati: Audiweb - Nielsen//NetRatings – NetView (Agosto 2003)

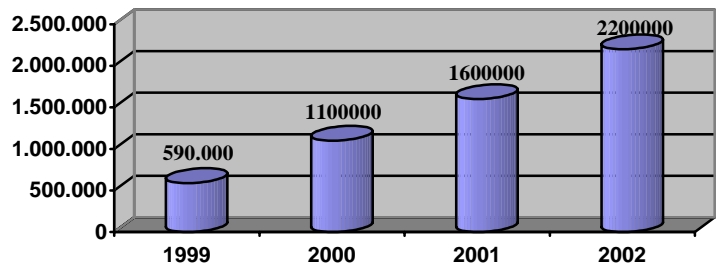

**Figura 4 -** Penetrazione Internet: PMI connesse a Internet (Assinform, 2001)

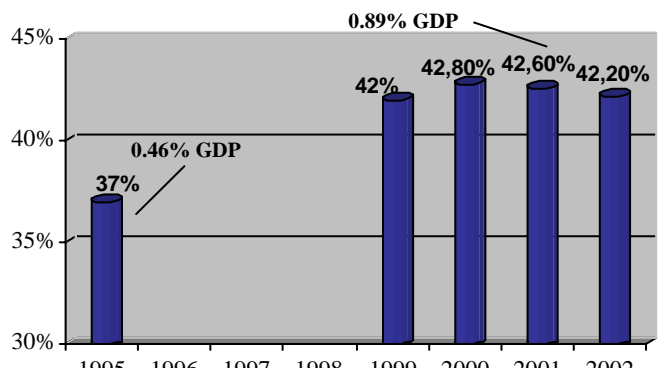

**Figura 5 -** Peso delle PMI sul mercato IT (percentuale) (Assinform, 2003)

**Tabella 2 -** Investimenti IT nel 2000-2002 (milioni diEuro) in Italia: per regione e area geografica (Assinform, 2003a)

| REGIONI | 2000 | 2001 | Δ 2001/00 | 2002 | Δ 2002/01 |
|---|---|---|---|---|---|
| Lombardia | 4.190,6 | 4.525,1 | 8,0% | 4.550,5 | 0,6% |
| Piemonte | 2.182,4 | 2.348,9 | 7,6% | 2.252,5 | -4,1% |
| Liguria | 552,9 | 593,2 | 7,3% | 547,2 | -7,7% |
| Valle d'Aosta | 60,1 | 64,9 | 7,9% | 60,2 | -7,1% |
| **Nord Ovest** | **6.986,0** | **7.532,0** | **7,8%** | **7.410,4** | **-1,6%** |
| Veneto | 1.553,5 | 1.708,1 | 9,9% | 1.730,3 | 1,3% |
| Trentino A. A. | 413,2 | 438,6 | 6,1% | 425,2 | -3,1% |
| Friuli V.G. | 415,2 | 456,5 | 9,9% | 445,2 | -2,5% |
| Emilia R. | 1.547,1 | 1.671,6 | 8,0% | 1.600,2 | -4,3% |
| **Nord Est** | **3.929,0** | **4.274,8** | **8,8%** | **4.200,9** | **-1,7%** |
| Toscana | 1.152,2 | 1.266,7 | 9,9% | 1.245,1 | -1,7% |
| Marche | 450,9 | 461,6 | 2,4% | 466,0 | 1,0% |
| Umbria | 248,2 | 264,4 | 6,5% | 263,3 | -0,4% |
| Lazio | 3.644,8 | 3.921,4 | 7,6% | 3.806,7 | -2,9% |
| **Centro** | **5.496,0** | **5.914,1** | **7,6%** | **5.781,1** | **-2,3%** |
| Campania | 872,2 | 956,3 | 9,6% | 910,1 | -4,8% |
| Abruzzo | 177,1 | 194,1 | 9,6% | 185,9 | -4,2% |
| Puglia | 488,5 | 530,8 | 8,7% | 514,7 | -3,0% |
| Molise | 47,5 | 48,6 | 2,4% | 45,6 | -6,2% |
| Basilicata | 84,6 | 86,6 | 2,4% | 80,6 | -6,9% |
| Calabria | 203,8 | 218,0 | 6,9% | 209,2 | -4,0% |
| Sicilia | 500,9 | 550,7 | 9,9% | 534,3 | -3,0% |
| Sardegna | 173,6 | 172,0 | -0,9% | 162,9 | -5,3% |
| **Sud & Isole** | **2.548,0** | **2.757,0** | **8,2%** | **2.643,4** | **-4,1%** |
| **Totale Italia** | **18.959,0** | **20.478,0** | **8,0%** | **20.035,8** | **-2,2%** |

**Tabella 3 -** Spesa IT per lavoratore nelle regioni italiane (Assinform, 2003a)

| REGIONI | Spesa IT per dipendente (€) | REGIONI | Spesa IT per dipendente (€) |
|---|---|---|---|
| Lazio | 1.686 | Umbria | 749 |
| Piemonte | 1.195 | Marche | 737 |
| Valle d'Aosta | 1.091 | Campania | 536 |
| Lombardia | 1.053 | Basilicata | 445 |
| Trentino A.A. | 998 | Molise | 430 |
| Friuli V.G. | 854 | Puglia | 427 |
| Veneto | 844 | Abruzzo | 399 |
| Liguria | 844 | Calabria | 389 |
| Emilia R. | 841 | Sicilia | 386 |
| Toscana | 774 | Sardegna | 302 |

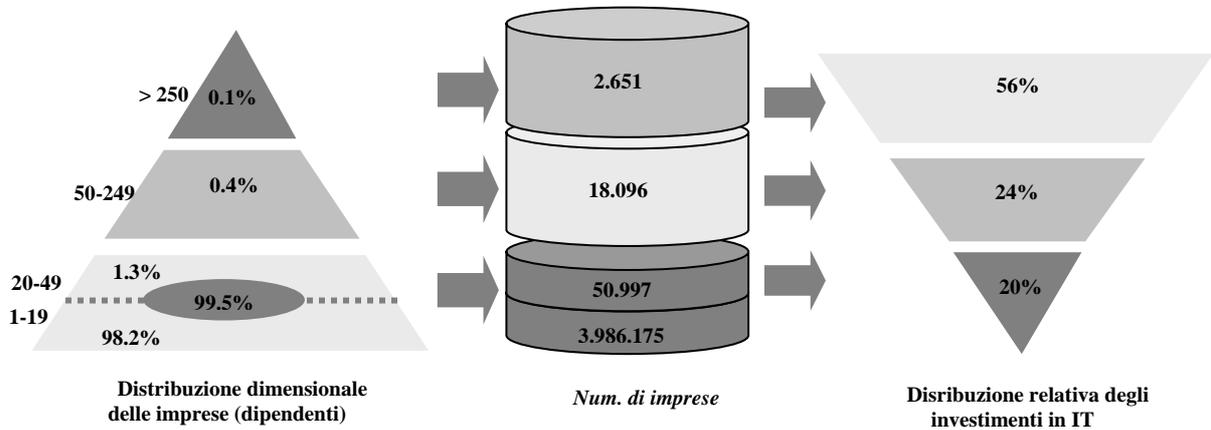

**Figura 6 –** Spesa IT per classi dimensionali in Italia (2002) (Koch, 2003)

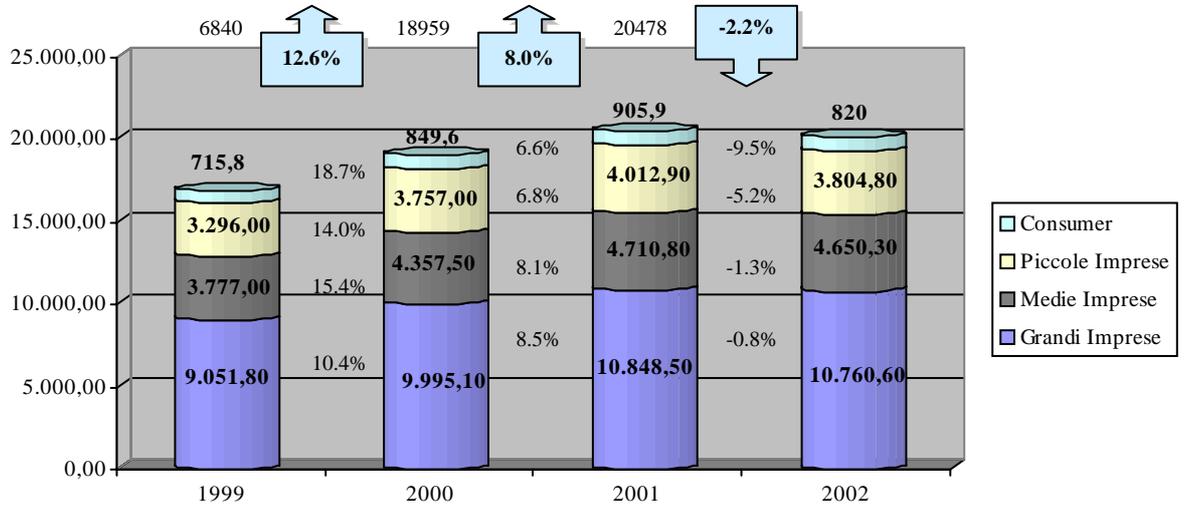

**Figura 7 -** Crescita del mercato IT italiano in relazione alle dimensioni delle imprese (2000-2002) (Capitani, 2003)

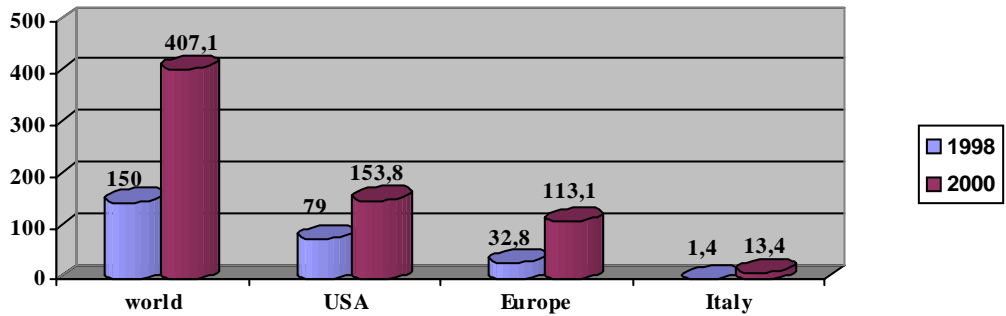

**Figura 8 -** Numero (milioni) di utenti Internet (almeno 1 accesso nei tre mesi precedenti l'indagine) (Assinform, 2001)

### 3.3 Ragioni della PMI italiana per la l'adozione e la non-adozione delle ICT.

Il "cattivo" uso delle ICT da parte delle PMI italiane continua ad apparire come un paradosso: da una parte abbondano gli indici relativi all'importanza economica delle PMI (e molti tra essi sono riconosciuti come "buoni": impatto delle PMI sul PIL, export, occupazione, ecc.), dall'altra, le indagini quantitative descrivono un quadro di PMI mal-equipaggiate e arrancanti sul fronte dello sfruttamento delle ICT a supporto del vantaggio competitivo. Ma quali sono le ragioni che spingono le PMI del nostro Paese a mantenere soluzioni IT nulle od obsolete? Tentiamo di fornire una risposta a tale quesito attraverso le evidenze emerse da alcune indagini effettuate da Camere di Commercio e ricerche universitarie.

#### 3.3.1 Non adozione delle ICT: skill shortage
In uno studio del 2000, Formaper (azienda speciale della CCIAA di Milano) ha analizzato un campione di 200 PMI site in provincia di Milano con lo scopo di comprendere come esse strutturino il processo di creazione della domanda di ICT.
Le imprese sono state segmentate sulla base della presenza di una rete e della connessione Internet (Fig.9): i risultati dell'indagine mostrano che esiste una relazione diretta tra le dimensioni delle imprese e l'intensità nell'uso delle ICT, anche quando le imprese considerate sono mediamente molto piccole. Non solo, dallo studio emerge anche che le difficoltà sperimentate dalle PMI nell'adottare soluzioni ICT innovative dipendono principalmente dalla carenza di competenze (*skill shortage*) adeguate ad effettuare le necessarie valutazioni, anche nel caso delle imprese maggiormente "innovative" e reattive (cfr. Fig.10), impedendo di fatto alle imprese di sviluppare una visione strategica sufficientemente lungimirante in grado di produrre efficaci investimenti in tecnologia. Questa evidenza empirica è consistente con la tipologia di ostacoli all'adozione di ICT evidenziata in letteratura (cfr. pag.10).

**Network e/o sistema centrale**

|  | NO | SI |
|---|---|---|
| **Internet NO** | GRUPPO I: non sistemiche<br>• <10 dipendenti<br>• Settori tradizionali<br>• Una sede<br>• Proprietà familiare<br>• Fondate anni '60-'70<br>• Nessun sistema di certificazione della qualità | GRUPPO II: sistemiche isolate<br>• 15-24 dipendenti<br>• Lavorazioni metalliche e meccanica di precisione<br>• Misto tra controllo familiare e non fondate prima '60<br>• sistema di certificazione della qualità<br>• export<br>• innovazione di prodotto e di processo<br>• espansione verso nuovi mercati |
| **Internet SI** | • mercato interno<br>• innovazione di processo e/o ricerca di nuovi mercati<br>• margini risicati | GRUPPO III: sistemiche connesse<br>• 25 dipendenti<br>• Servizi alla produzione, industria libraria<br>• più sedi nella maggioranza dei casi<br>• controllo spesso suddiviso<br>• alti volume di export<br>• innovazione di processo, organizzativa, strategica, marchi e brevetti<br>• margini elevati |

**Figura 9** – Segmentazione delle PMI analizzate (elaborazione su dati Formaper)

| Gruppo | Decisore d'acquisto | Dimensione domanda ICTs [1] | Equipaggiamento informatico e attività informatizzate [1] | numero fornitori | Criterio selezione fornitori | Principali limiti per domanda di ICT |
|---|---|---|---|---|---|---|
| I | Basso titolo di studio, proprietario/parente, 40-50 anni, background umanistico, funzioni EDP/amministrative | bassa | bassa | 1 | prezzo | Scarse conoscenze tecniche, scarsa percezione dei servizi e delle loro utilità |
| II | Cultura media, quadro, 30-39 anni, background tecnico, funzioni EDP/amministrative | media | medio | 2-3 | prezzo, servizio, comunicazione | Scarse conoscenze tecniche, resistenze culturali, prezzo |
| III | Elevato titolo di studio, manager, <30 anni, background ingegneristico, funzioni EDP/amministrative | medio-alta | medio-alto | >3 | servizio, comunicazione, qualità | Scarse conoscenze tecniche |

[1] relativa al campione considerato

**Figura 10** – Elementi caratterizzanti il processo di acquisto di ICT nelle PMI (elaborazione su dati Formaper)

### 3.3.2 ICT e distretti: un problema di visione strategica

È altrettanto interessante esaminare i dati emersi dal lavoro svolto nel 2002 da Brunetti, Micelli and Minoja sul problema dell'adozione delle ICT nei distretti industriali italiani del nord-est. I tre ricercatori hanno analizzato un totale di 311 PMI appartenenti a distretti situati in Lombardia e nel Veneto, principalmente con lo scopo di verificare se sia possibile individuare comportamenti comuni nel processo di adozione di ICT tra imprese differenti appartenenti a settori differenti e site in contesti differenti.

Un primo risultato - abbastanza sorprendente - è riassunto in Tab.4: la maggior parte delle imprese, anche se impegnate in attività non direttamente correlate con l'ICT (appartengono tutte a settori produttivi tradizionali: mobile, meccanica, tessile) dimostra un elevato grado di informatizzazione. Ciò significa che il PC è finalmente uscito dal ristretto ambito delle funzioni amministrative. Ma, a un'analisi appena un po' più approfondita, appare evidente che il PC è in realtà vissuto alla stregua di una *commodity*, utile principalmente per consultare l'e-mail e accedere ai servizi di *banking online* (cfr. Tab.5 e 6). Cionondimeno è interessante sottolineare che, probabilmente anche grazie alla "sensibilizzazione" derivante dalla capillare diffusione del telefono cellulare in Italia, le imprese intervistate hanno dichiarato un forte interesse per la costruzione di reti basate su tecnologie *wireless* (meno costose e più semplici da installare e mantenere, e maggiormente flessibili).

Riguardo alle differenze tra le due macro-aree geografiche (Lombardia e Veneto) si rileva che:
− la distribuzione percentuale delle differenti tecnologie considerate nell'indagine (cfr. Tab.5) è simile;
− mentre una percentuale sostanziosa (23% and 24% rispettivamente) delle imprese di entrambe le aree esternalizza completamente la funzione informatica, una percentuale maggiore di imprese lombarde (31,1% vs. 20,9% delle venete) ritiene importante detenere un controllo totalmente interno su questa funzione strategica.

Dall'analisi dell'uso del sito web (cfr. Tab. 7) provengono altri interessanti spunti di riflessione. Ad esempio si constata che, malgrado rispettivamente il 68,9% e l'85,5% delle imprese lombarde e venete dichiari di possedere un sito web, l'*e-commerce* occupa saldamente la penultima posizione nella classifica dell'adozione delle tecnologie (cfr. Tab. 7). Inoltre, nella maggior parte dei casi il sito web in questione è una sorta di vetrina online dell'impresa, il cui scopo principale è fornire informazioni sui prodotti e i servizi. Tuttavia molte imprese ritengono che sia necessario un miglioramento sia dei contenuti che della presentazione grafica del sito, e hanno previsto delle opportune voci di spesa. Riguardo alle possibili evoluzioni del sito (le sue migliorie occupano il primo posto nella classifica delle priorità del 63,5% delle imprese lombarde e del 72% di quelle Venete – cfr. Tab. 8), si ritiene importante soprattutto migliorare il supporto all'interattività con i clienti. È anche interessante notare che, mentre una delle principali conseguenze del sito web è l'incremento nel numero dei clienti (rispettivamente pari al 29.6% e al 32.5% per le imprese lombarde e venete), il principale motivo per non adottare soluzioni di *e-commerce* è l'inadeguatezza del canale Internet (53% e 62.3% delle risposte). Secondo Brunetti,

Micelli e Minoja, sembra che la percezione di inadeguatezza del canale Internet per la vendita *online* derivi dalla preoccupazione di creare sovrapposizioni e frizioni con il canale distributivo tradizionale, mentre gli aspetti legati alla sicurezza non sono più "sentiti" come in passato. In ogni caso la risposta alla domanda "perché ritiene Internet inadeguata?" è quasi sempre molto generica, denotando così una lacunosa conoscenza delle potenzialità del mezzo: ad esempio il 66% e il 72,2% rispettivamente delle imprese lombarde e venete non sono a conoscenza dell'esistenza dei *marketplace*. Ciononostante, in qualche modo le potenzialità strategiche delle ICT sono percepite, visto che l'87,5% e l'82,2% delle imprese ritiene che l'uso delle ICT possa avere un'importanza strategica medio-alta. Tuttavia, malgrado le priorità dichiarate (cfr. Tab. 8), il budget allocato in realtà agli investimenti in ICT è molto contenuto (meno dell'1% del fatturato è riservato a questo tipo di investimenti rispettivamente dal 47,2% e dal 57,3% delle imprese delle due aree).

**Tabella 4** – imprese classificate per classe di informatizzazione (% num. PC/lavoratore) (Brunetti, Micelli & Minoja, 2002)

| Classe di informatizzazione | Lombardia | Veneto |
|---|---|---|
| 0 – 25% | 6.7 | 1.0 |
| 25.1 – 50% | 8.7 | 4.8 |
| 50.1 – 100% | 38.5 | 27.6 |
| Più del 100% | **46.2** | **66.7** |
| Total | 100 | 100 |

**Tabella 5** – diffusione delle ICT nei distretti industriali della Lombardia e del Veneto (percentuale) (Brunetti, Micelli & Minoja, 2002)

| tecnologia | Lombardia | Veneto |
|---|---|---|
| e-mail | **96.2** | **97.3** |
| ISDN | 84.9 | 90.0 |
| Corporate banking | 73.6 | 83.6 |
| Web site | 68.9 | 85.5 |
| Mobile telephones network | 37.7 | 48.2 |
| ERP | 18.9 | 17.3 |
| ADSL | 16.0 | 12.7 |
| EDI | 11.3 | 9.1 |
| Groupware | 9.4 | 10.0 |
| e-commerce | 4.7 | 3.6 |
| Videoconferencing | 3.8 | 5.5 |

**Tabella 6** – dotazione interna di e-mail (percentuale) (Brunetti, Micelli & Minoja, 2002)

| e-mail | Lombardia | Veneto |
|---|---|---|
| Ogni ufficio | **51.0** | **50.5** |
| Alcuni uffici | 42.2 | 33.6 |
| Una sola mailbox | 6.9 | 15.9 |

**Tabella 7 –** adozione sito web ed e-commerce
(elaborazione su dati di Brunetti, Micelli & Minoja)

| | LOMBARDIA | | VENETO | |
|---|---|---|---|---|
| **Funzioni presenti e future del sito Web (%)*** | | | | |
| | **Presenti** | **Future** | **Presenti** | **Future** |
| Presentazione azienda | **97.3** | **50.0** | **97.8** | 48.9 |
| Informazioni sui prodotti | 87.7 | **50.0** | 87.1 | **55.4** |
| Presentazione catalogo | 45.2 | 42.5 | 59.1 | 45.7 |
| Supporto clienti | 16.7 | 26.0 | 21.5 | 35.2 |
| Raccolta informazioni dai clienti | 20.5 | 20.5 | 26.9 | 28.3 |
| Test prodotti | 8.2 | 9.6 | 8.6 | 14.1 |
| Vendite retail | 6.8 | 17.8 | 4.3 | 14.1 |

* percentuale calcolata sulle imprese che dichiarano di voler ulteriormente investire sul sito web

| **Investimenti futuri sul sito web (%)** | | |
|---|---|---|
| Miglioramento qualità grafica | 56.8 | 67.2 |
| Miglioramento qualità contenuti | **90.9** | **89.7** |
| Intergrazione coi processi | 25.0 | 46.6 |
| Integraz.con il sistema del valore | 50.0 | 63.8 |

| **Principali risultati dovuti al sito web (%)°** | | |
|---|---|---|
| Aumento umero clienti | 29.6 | 32.5 |
| Aumento vendite | 14.8 | 15.6 |
| Aum. vendite prodotti di nicchia | 16.7 | 11.7 |
| Aumento interazione coi clienti | **50.0** | **61.0** |
| Suggerimenti sui prodotti | 16.7 | 26.0 |
| Maggiori feedback dal mercato | 16.7 | 32.5 |
| altro (promozione) | 27.8 | 19.5 |

° percentuale calcolata sulle imprese che dichiarano impatto non nullo del sito web

| **Cause per la non-adozione dell'e-commerce (%)** | | |
|---|---|---|
| Canale inadeguato | **53.0** | **62.3** |
| Frizini con la rete distributiva | 10.0 | 12.3 |
| Progetto in via di valutazione | 9.0 | 13.2 |
| Mancanza di risorse | 9.0 | 5.7 |
| Sicurezza | 0.0 | 6.6 |
| Progetto in corso | 5.0 | 5.7 |
| Costi | 3.0 | 1.9 |
| Altro | 24.0 | 15.0 |

**Tabella 8 –** investimenti futuri in ICT
(elaborazione su dati di Brunetti, Micelli & Minoja, 2002)

| | LOMBARDIA | VENETO |
|---|---|---|
| **Impatto sul management delgli investimenti in ICT (%)** | | |
| Trasformazione dell'attività | 6.7 | 8.3 |
| alto | 42.3 | 33.3 |
| marginale | **48.1** | **50.9** |
| nullo | 2.9 | 7.4 |
| **Priorità tecnologiche (investimenti futuri) (%)** | | |
| rinnovo diLAN e hardware | 48.1 | 57.9 |
| Acquisizione di un ERP | 15.4 | 20.4 |
| Sviluppo soluzione di Groupware | 11.7 | 19.4 |
| Arricchimento sito Web | **63.5** | **72.0** |
| Soluzioni integrate con i fornitori | 25.0 | 38.9 |
| Soluz. integrate con la strut. commerciale | 45.2 | 58.3 |
| **Attore promotore dell'adozione di ICT (%)** | | |
| imprenditore | **61.3** | 42.6 |
| Responsabile del sistema informativo | 57.5 | **67.6** |
| Consulenti esterni | 22.6 | 25.9 |
| fornitori | 3.8 | 3.7 |
| clienti | 5.7 | 5.6 |
| associazioni, ecc. | 4.7 | 1.9 |

### 3.3.3 Non-adozione di soluzioni ICT: inadeguatezza dell'offerta

Altre interessanti suggestioni circa le cause profonde della scarsa o non sufficientemente efficace adozione delle ICT da parte della PMI italiana, derivano dai risultati di una ricerca effettuata su di un campione di 1013 piccole e medie imprese localizzate nel Friuli-Venezia Giulia (Ronzini, 2001). Da questi risultati scopriamo, ad esempio, che la principale ragione per la quale l'azienda non viene dotata di connessione Internet è la "mancanza di tempo" (cfr. Fig. 11), mentre, però, Internet viene percepita come "un'opportunità" dall'80% del campione (cfr. fig. 14), e come uno strumento utile per "seguire l'evoluzione del mercato" (68,5%), "migliorare il servizio al cliente" (61%) e "fidelizzare i partner" (57%).

Nuovamente qualcosa sembra mancare nel puzzle complessivo, dato che le ragioni per utilizzare soluzioni basate su tecnologie Internet, sommate alla percezione positiva che se ne ha, sembrerebbero più che sufficienti per decidere di provare a superare le barriere derivanti dai vincoli di tempo. Un indizio a tal proposito può essere derivato dai dati riportati in Fig. 13, che ci dicono implicitamente che la mancanza di competenze interne (*skill shortage*) – e non solo di tempo! – e le difficoltà nella valutazione dell'investimento sono tra le principali cause, ad esempio, per non sviluppare (ulteriormente) il sito web.

Non solo, Ronzini procede nella sua analisi considerando anche il lato offerta di soluzioni tecnologiche per le PMI (che si concretizza principalmente in software house, *Internet Service Providers* - ISP, *Applications Service Providers* – ASP), e scopre un'ulteriore fattore di osacolo all'adozione di ICT:

- <u>software house</u>: generalmente sono il primo interlocutore per l'impresa che desidera sviluppare una soluzione ICT o web, ma in genere esse sono quasi totalmente prive di "*marketing sense*", e si limitano, anziché a conferire la consulenza che sarebbe necessaria a chi si affaccia su un media nuovo, a implementare le idee dell'EDP manager (laddove esiste) dell'impresa. Ovviamente questo approccio può difficilmente garantire risultati soddisfacenti.

- <u>ISP</u>: il loro core business è la fornitura di connettività; essi, quindi, tendono a sottovalutare gli aspetti relative all'integrazione di soluzioni innovative Internet-based con il sistema informativo già in essere presso i loro clienti. Nuovamente un simile approccio non può portare molto lontano.

- <u>ASP</u>: offrono servizi ad alto valore aggiunto e di alta qualità, generalmente in partnership con imprese di telecomunicazioni, società marketing, ecc. In genere la loro offerta si concentra soprattutto sulle grandi città, e ha costi che praticamente nessuna PMI può sognare di permettersi.

Pertanto, anche se l'approccio dei fornitori si sta gradualmente spostando verso obiettivi convergenti con quelli della PMI (cfr. Fig. 15), le conclusioni tratte da Ronzini sembrano essere ancora valide: l'ICT è di fatto uno strumento capace di supportare in maniera appropriata l'efficienza e l'efficacia della PMI, ma, nella maggioranza dei casi, le sue potenzialità sono sfruttate in maniera molto superficiale, è quindi necessario investire in *educazione*, sottolineando la necessità di adattare le soluzioni ICT ai bisogni reali della PMI, e non viceversa (come troppo spesso continua ad accadere).

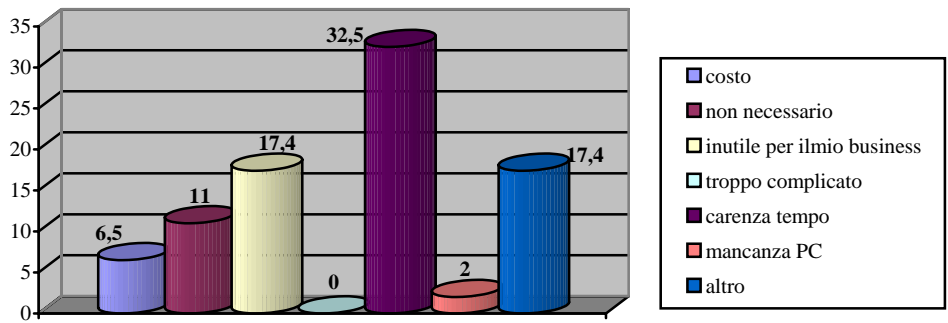

**Figura 11 –** Ragioni per non connettersi ad Internet (%)

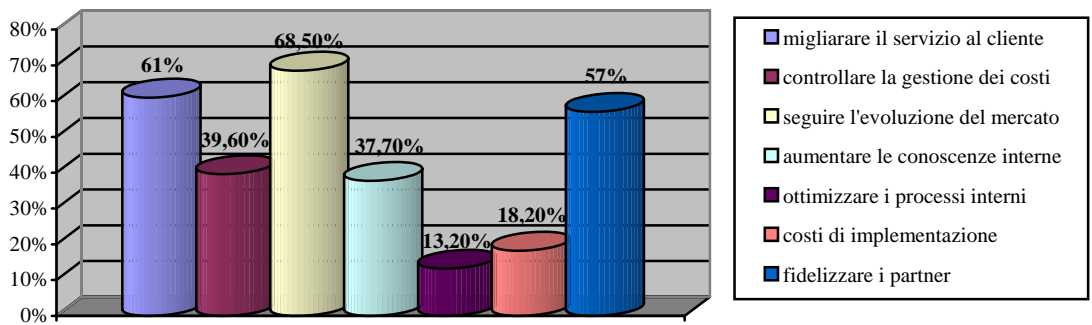

**Figura 12 -** Ragioni per l'uso di Internet (%)

Il rapporto tra ICT e PMI italiane e le problematiche economico-organizzative dell'OS.    23

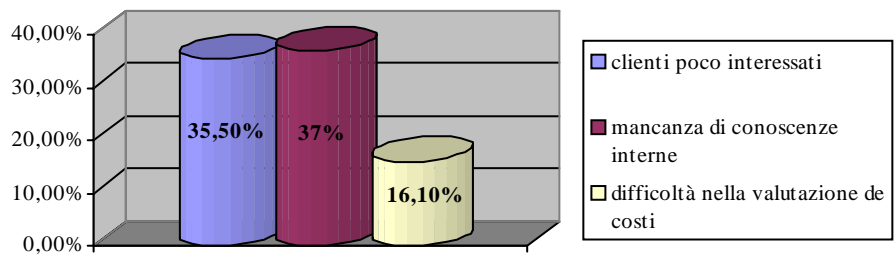

**Figura 13 -** Ostacoli allo sviluppo del sito

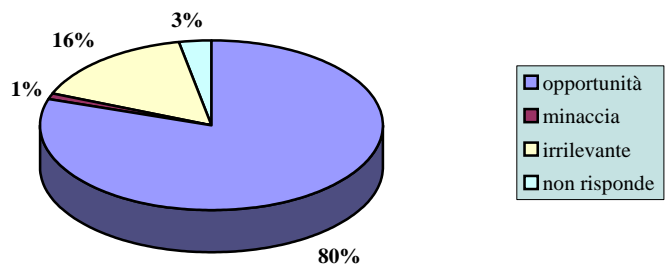

**Figura 14 -** Come le PMI percepiscono Internet

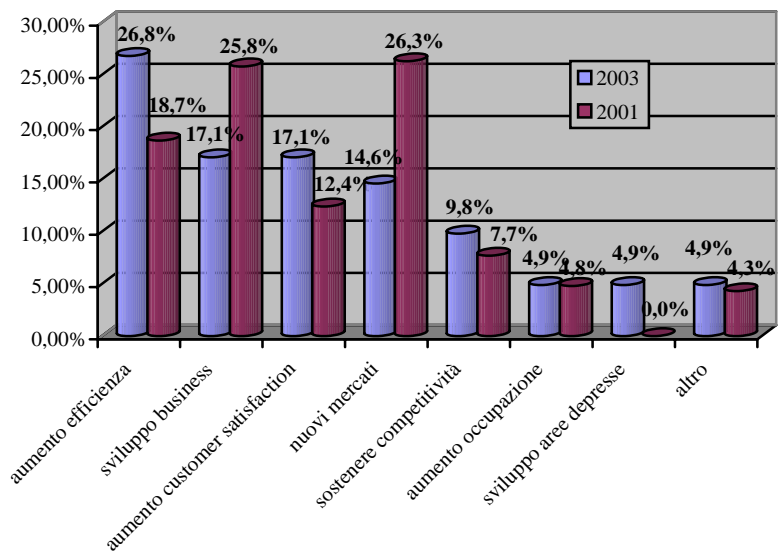

**Figura 15 -** obiettivi delle soluzioni ICT secondo l'offerta (Assinform, 2003)

## 4. PMI e investimenti in ICT.

Trovata conferma empirica della tipologia delle difficoltà nell'adozione di soluzioni ICT con valenza strategica nella PMI, è ora interessante approfondire ulteriormente l'analisi, allo scopo di comprendere quali siano le strategie di investimento adottate dalle PMI italiane per le ICT.

A tal fine è interessante analizzare criticamente i dati che emergono dall'ormai annuale *survey* sullo stato dell'adozione delle ICT nelle PMI effettuata dall' "Osservatorio ICT nelle PMI" della School of Management del Politecnico di Milano. In particolare l'indagine 2006 ha visto il coinvolgimento di due differenti campioni (entrambi statisticamente significativi) di imprese (cfr. Tab. 9): il primo campione – trasversale a diversi settori – è stato utilizzato per indagare quali siano le aree di investimento prioritarie per le PMI e quali logiche esse adottino per finanziare tali investimenti, mentre il secondo campione – focalizzato sul manifatturiero – ha fornito informazioni riguardo al grado di penetrazione delle ICT e al relativo livello di spesa.

**Tabella 9 -** Finalità e campioni dell'indagine 2006 della School of Management del Politecnico di Milano

| Obiettivo indagine | Campione | Fascia dimensionale | Settore |
|---|---|---|---|
| - aree di investimento<br>- logiche di finanziamento | 500 | 2 – 500 dipendenti | - manifatturiero<br>- costruzioni<br>- commercio<br>- servizi (no finanziari) |
| - livello diffusione ICT<br>- livello di spesa | 646 | 10 – 500 dipendenti | - manifatturiero |

Ovviamente non tutti i dati estrapolabili dalla *survey* sono utili per le finalità di questo lavoro. Nel seguito, pertanto, riassumeremo solo le osservazioni maggiormente rilevanti.

### 4.1 Come investono le PMI?

L'indagine del Politecnico di Milano pone in luce le dinamiche di investimento delle PMI tra 2 e 500 dipendenti (1° campione). Alle imprese è stato chiesto se avessero intenzione di investire nel corso del 2006 in progetti innovativi in almeno una delle seguenti aree:

- innovazione di processo
- innovazione di prodotto
- ICT
- valorizzazione del brand
- internazionalizzazione

Per il nostro lavoro sono rilevanti in particolare le informazioni riguardanti gli investimenti in ICT, tuttavia non è possibile estrapolare dati sufficientemente puntuali da quelli pubblicati nella *survey*: in particolare non è possibile incrociare la classe dimensionale delle imprese con la propensione all'investimento in ICT. Il *report* della *survey,* infatti, riporta abbastanza genericamente che la propensione all'investimento delle imprese di dimensioni minori è molto più contenuta rispetto a quelle di dimensioni più considerevoli. È inoltre da osservare che, presumibilmente, la maggioranza degli investimenti in innovazione che ricadono in tutte le classi analizzate (innovazione di processo, di prodotto, internazionalizzazione, ecc.) implichino in maniera più o meno diretta anche un investimento in ICT.

**Tabella 10** - Come le PMI pianificano gli investimento per il 2006

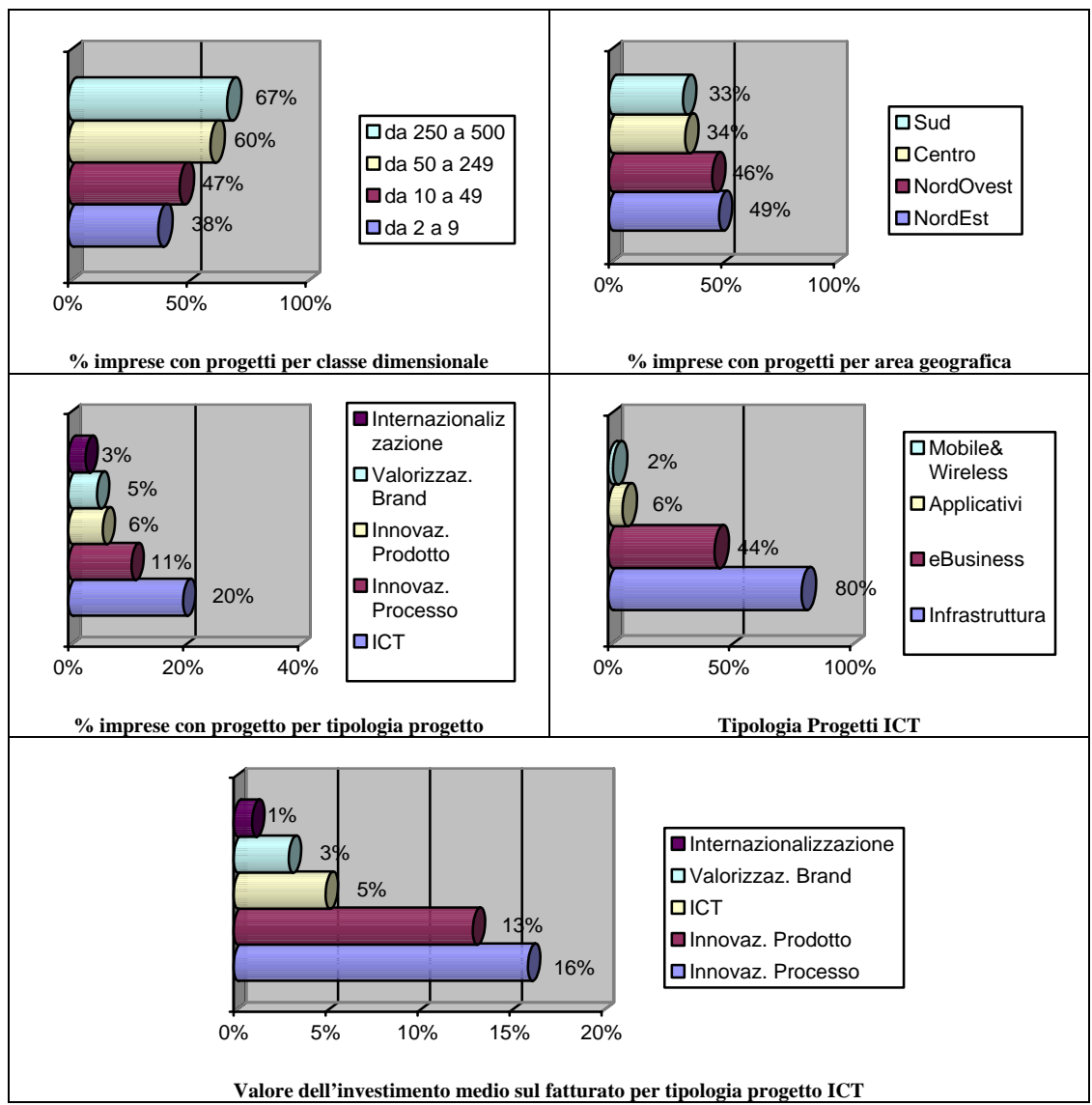

In Tab. 10 sono riportati i dati aggregati riguardanti la percentuale di imprese intervistate per *classe dimensionale*, la percentuale di imprese che progettano di *investire in almeno una delle aree* considerate disaggregate per area geografica, la percentuale di imprese che intende investire per *tipologia di progetto* (i progetti considerati sono naturalmente relativi alle aree di investimento prese in esame) e l'entità media dell'investimento. I dati, inoltre, sono disaggregati per quanto riguarda gli *investimenti in ICT*, specificando la tipologia di investimento per macro-aree: mobile&wirelss, applicativi, sviluppo di soluzioni di eBusiness, potenziamento dell'infrastruttura.

La ricerca riferisce inoltre lo scarso ricorso ai mezzi di finanziamento per sostenere progetti nell'area ICT. Ciò sembra implicare una percezione di scarso impatto strategico di quest'area. Come risulta evidente dalla precedente tabella, infatti, in genere gli investimenti sono di entità piuttosto contenuta e riguardano innovazioni incrementali autofinanziate. Questo atteggiamento risulta maggiormente frequente nelle imprese di dimensioni minori.

È inoltre interessante osservare la correlazione che lega il posizionamento strategico delle imprese e la loro propensione ad investire (cfr. Fig. 16): purtroppo tali dati non sono disaggregati nelle differenti tipologie di investimento, non è perciò possibile comprendere quanto tali investimenti impattino sulle ICT. Sarebbe anche interessante poter disporre dal dato disaggregato sulla base del settore di appartenenza delle imprese, sulla classe dimensionale e sulla loro "anzianità", per ogni quadrante della figura: ciò consentirebbe di verificare se esista – come verosimilmente si può immaginare – un'ulteriore correlazione della propensione ad investire anche con queste tre variabili.

| Attrattività business + | 57% imprese che investono<br>18% rapporto medio investimento su fatturato | 72% imprese che investono<br>19% rapporto medio investimento su fatturato |
|---|---|---|
| Attrattività business - | 36% imprese che investono<br>9% rapporto medio investimento su fatturato | 57% imprese che investono<br>12% rapporto medio investimento su fatturato |
| | - Posizione competitiva + | |

**Figura 16** - correlazione tra posizionamento competitivo e investimenti

### 4.2 Quali sono le dotazioni tecnologiche infrastrutturali delle PMI?

Sempre dalla stessa indagine è possibile estrapolare alcune informazioni di scenario sulla dotazione tecnologica delle PMI italiane. A livello *infrastrutturale*, dal campione esaminato sembra emergere una relazione tra il numero di PC o portatili per dipendente e il settore nel quale l'impresa si trova ad operare (cfr. Fig. 17).

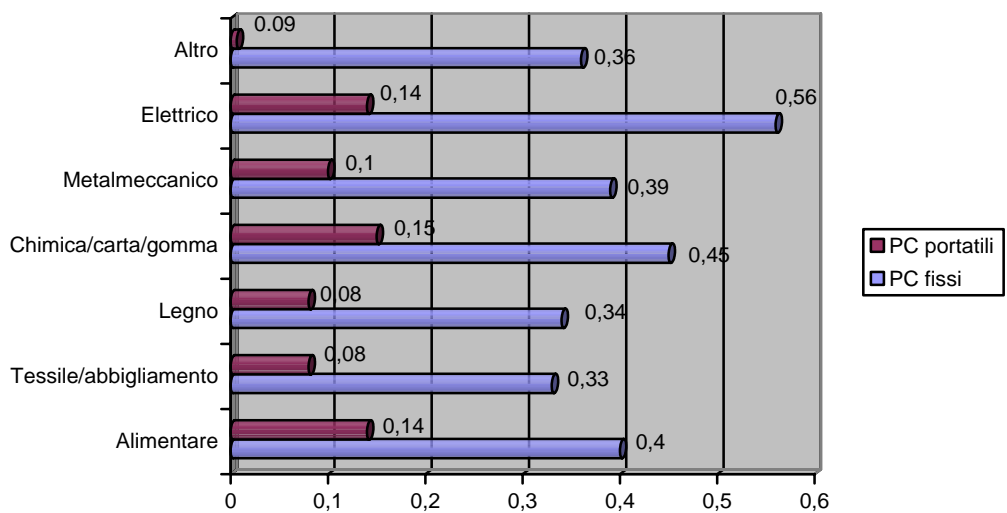

**Figura 17** - Numero medio di PC per dipendente in diversi settori del manifatturiero

Per quanto riguarda, invece, la tipologia di *sistemi operativi impiegati sui PC* nelle PMI – dato rilevante per lo scopo di questo lavoro – si osserva una nettissima prevalenza dei sistemi Microsoft (Windows XP, 2000, 98, NT), che complessivamente sono installati sul 97,6% del parco macchine (cfr. Fig. 18). Leggermente più variegata è la situazione dei *sistemi operativi installati sui server aziendali*, dove complessivamente i sistemi Microsoft si attestano attorno al 66% del totale (cfr. Fig. 19). Simile è la situazione se consideriamo i *sistemi operativi del server sui quali risiede l'applicativo gestionale* delle imprese: notiamo infatti che i sistemi Microsoft vengono utilizzati sul 65% delle macchine (cfr. Fig. 20). Da notarsi, inoltre, che in un 12% dei casi il gestionale risiede su PC dotati di Windows XP: si tratta prevalentemente di imprese di dimensioni molto ridotte, per le quali una soluzione basata su server potrebbe risultare sovradimensionata.

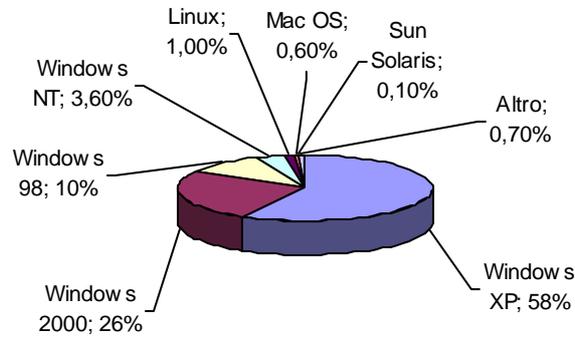

**Figura 18 -** Sistemi operativi dei PC aziendali

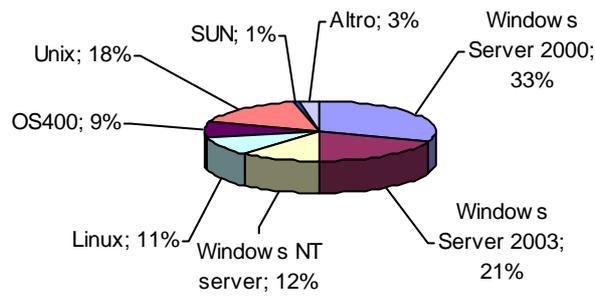

**Figura 19** - Sistemi operativi del server aziendali

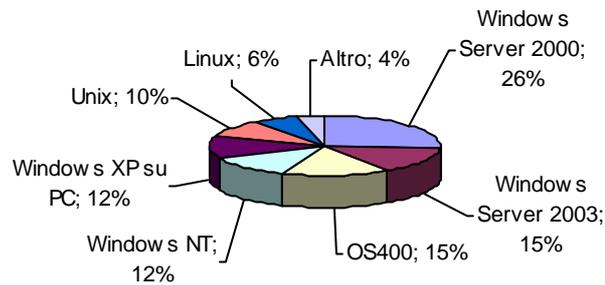

**Figura 20** - Sistemi operativi dei server per il gestionale

Il rapporto tra ICT e PMI italiane e le problematiche economico-organizzative dell'OS. 29

**4.3 Quali sono le dotazioni applicative delle PMI nel manifatturiero?**

Sempre dall'indagine del Politecnico di Milano è possibile estrapolare alcuni interessanti dati riguardanti il tipo di applicativi utilizzati in prevalenze nelle PMI del settore manifatturiero (cioè: sistemi gestionali, CAD, PLM – Produc Lifecycle Management, eBusiness, applicazioni mobile&wirelss).

Da notare che il ricorso alle soluzioni basate su mobile&wireless appare in crescente aumento a causa del minore TCO (Total Cost of Ownership) e della maggiore versatilità rispetto a soluzioni basate su architetture "fisse".

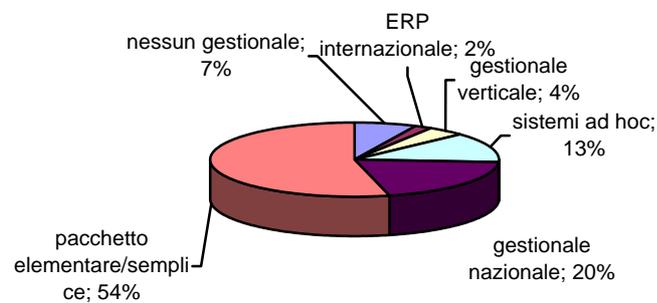

**Figura 21** - Tipologia dei gestionali impiegati nelle PMI

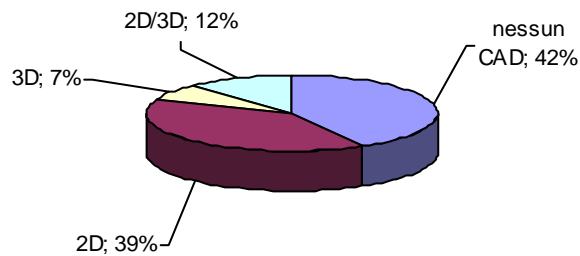

**Figura 22** - Diffusione dei differenti tipi di CAD nelle PMI

**4.4 Qual è la diffusione del software *open source* nelle PMI del manifatturiero?**

Un sezione del *report* particolarmente significativa per il nostro lavoro si focalizza sulla penetrazione del F/OSS nelle PMI del manifatturiero. Sebbene i dati riportati (cfr. Figg. 23 e 24) si limitino a illustrare la situazione del software infrastrutturale e applicativo, senza entrare in un maggior grado di dettaglio, è comunque interessante verificare che l'adozione di F/OSS:

− cresce proporzionalmente alla classe dimensionale
− è maggiormente accentuata nelle aree del nord Italia (soprattutto per quanto concerne gli applicativi)
− avviene particolarmente nel caso infrastrutturale.

È infine interessante ricordare che questi dati fanno riferimento a un campione di PMI statisticamente significativo, che si presume perciò contenere aziende appartenenti a tutte le fasce di posizionamento strategico, contrariamente a quanto avveniva in analoghe ricerche (seppur molto più dettagliate) effettuate recentemente, ad esempio dal TeDIS della Venice International University, ricerca dalla quale è tratta Fig. 25, che riporta la percentuale di diffusione dell'OS per classi di fatturato.

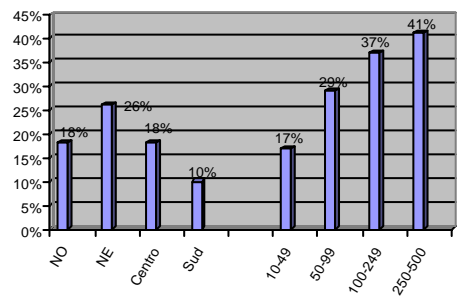

**Figura 23** - Diffusione F/OSS infrastrutturale (area geografica e numero dipendenti)

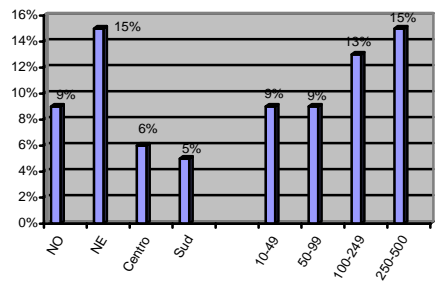

**Figura 24** - Diffusione F/OSS applicativo (area geografica e numero dipendenti)

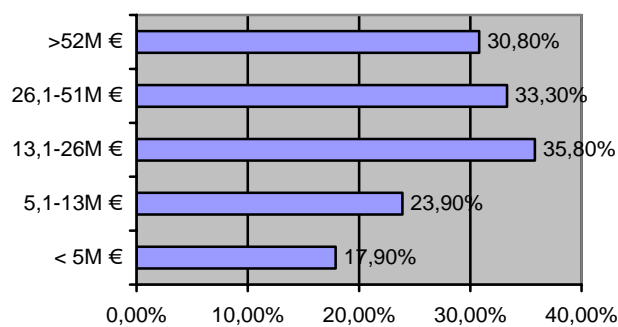

**Figura 25** - Diffusione % OS nelle imprese per classi di fatturato (milioni di €)

**4.5 Il ruolo del pivot dell'innovazione ICT**

L'indagine sullo stato di adozione delle ICT nelle PMI effettuata dal Politecnico di Milano nell'anno 2005 (cioè l'edizione precedente a quelle che abbiamo analizzato fino a questo punto), evidenzia alcune interessanti considerazioni relative a chi, in una PMI, svolga il ruolo di attore chiave nella decisione di investimento in ICT. Dato che questa informazione può risultare rilevante per un'azienda che si stia accingendo ad adottare un modello di business basato sul software OS, crediamo opportuno riportarla.

L'innovazione strategica efficace non può prescindere dalla compresenza di adeguate condizioni: *imprenditoriali*, *organizzative* e *gestionali*. A questo proposito l'analisi effettuata nell'indagine evidenzia tre prerequisiti fondamentali per la creazione di un substrato adeguato allo sviluppo di un atteggiamento propenso all'innovazione ICT in un'impresa:
− *committment del vertice strategico* (che nelle PMI può essere rappresentato dall'imprenditore stesso), che richiede una certa sensibilità nei confronti delle tematiche ICT;
− presenza in azienda di una figura professionale in grado di svolgere la funzione di *pivot ICT*, che implica la capacità di fungere da *liason* tra il vertice, eventuali funzioni tecniche, e fornitori esterni;
− un adeguato *presidio gestionale della funzione IT*. Ciò implica in alcuni casi la presenza della direzione IT, che spesso nelle PMI viene sostituita da un fornitore esterno che di fatto svolge il ruolo di partner strategico (per avere un'idea sintetica della diffusione della direzione IT nelle PMI, riportiamo i dati relativi al manifatturiero risultanti dall'indagine 2006 del Politecnico di Milano – cfr. Figg. 26, 27 e 28).

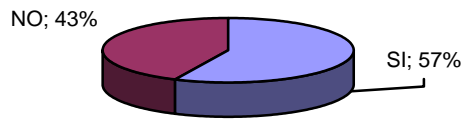

**Figura 26** - Presenza di una direzione IT nelle PMI del manifatturiero

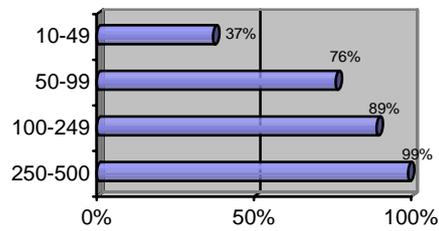

**Figura 27** - Presenza direzione IT per classe dimensionale

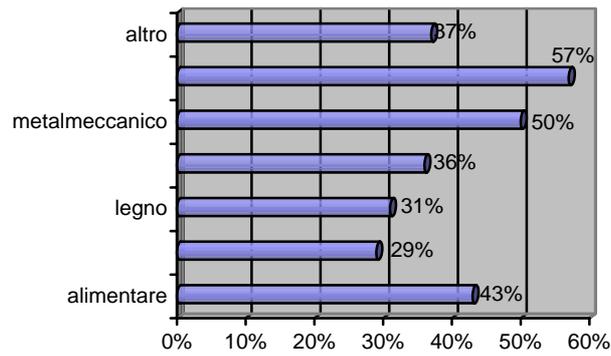

**Figura 28** - Presenza direzione IT per settore

Per lo scopo del presente lavoro è però interessante soffermarsi sulla modellizzazione delle caratteristiche della figura del Pivot ICT e del fornitore-partner effettuata dal Politecnico sulla base dei dati empirici raccolti nell'ambito dell'indagine 2005.

**4.5.1 Il Pivot ICT**
Il *pivot* ICT svolge un ruolo fondamentale nelle PMI (è colui che promuove l'innovazione ICT), è pertanto opportuno soffermarsi sui principali findings a questo proposito che emergono dalla ricerca effettuata dal Politecnico di Milano. Dall'analisi empirica è stata tratta una matrice (cfr. Fig. 29) che classifica i *pivot* delle imprese appartenenti al campione sulla base di due dimensioni: la loro provenienza (interna o esterna) e il ruolo ricoperto in azienda. I cerchi evidenziati in figura rappresentano il posizionamento del *pivot* rispetto agli assi nei 20 casi di studio che sono stati analizzati: come si vede la situazione è piuttosto eterogenea (ciò è in parte dovuto anche all'esiguità del campione). A questo proposito è essenziale ricordare che non tutte le direzioni IT sono in grado di svolgere adeguatamente il ruolo di *pivot* dell'innovazione ICT, dato che è necessario che dispongano di adeguate competenze, oltre che tecniche, nella gestione dei progetti (non dimentichiamo che un'innovazione ICT comporta nella maggior parte dei casi un impatto organizzativo anche potenzialmente rilevante, con la conseguente necessità di gestire in maniera appropriata "il cambiamento") e una buona capacità di "acquisto".

|  |  | **Provenienza interna** | | **Provenienza esterna** | |
|---|---|---|---|---|---|
|  |  | Imprenditore/figlio | Manager interno | Consulente | Manager di altra impresa |
| **Ruolo** | Amministratore delegato/Direttore Generale | ○ ○  ○ ○ | ○   ○ |  |  |
|  | Manager di linea | ○ | ○○○ ○ |  |  |
|  | Responsabile o manager direzione IT | ○ | ○○ ○ ○ ○ | ○ ○ | ○ |

**Figura 29** - Ruolo e provenienza del Pivot ICT

In tutti quei casi nei quali non è presente una direzione IT, il presidio gestionale dell'innovazione (cioè il ruolo di pivot) potrebbe essere efficacemente delegato a un partner esterno. Dato che, però, la figura che ricopre il ruolo di presidio gestionale dell'innovazione ICT è cruciale, è essenziale che – qualora questo ruolo sia riporto da un partner ICT esterno – esso sia dotato di adeguate:

- competenze tecniche sistemistiche e applicative
- competenze di system integration e correlati servizi (assistenza, help desk, ecc.)
- consulenza (analisi dei processi, ecc.)
- capacità di svolgere il ruolo di partner a medio-lungo termine

È evidente che tra le differenti tipologie di fornitore identificate sempre nell'ambito della ricerca 2005 effettuata dal Politecnico di Milano (cfr. Fig. 30), quella che maggiormente si adatta a questo ruolo di *partner ICT a tutto tondo* è il **system integrator**. In realtà, però, i fornitori in grado di svolgere questo ruolo nel mercato delle PMI sono per il momento ancora in numero esiguo.

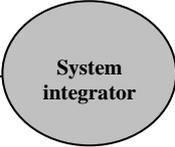

**Figura 30** - Tipologie di fornitori ICT

## 5. Conclusioni

Volendo riassumere i principali punti di attenzione, criticità e anche opportunità fin qui emersi, è bene distinguere i differenti piani di analisi nei quali è necessario articolare il processo di esame del complicato rapporto che le lega la piccola e media impresa italiana agli investimenti in ICT.

Come prima cosa, infatti, è importante distinguere la macrocategoria di tecnologie digitali alle quali si vuole fare riferimento: *ICT o semplicemente IT*. Diverso è infatti l'impatto organizzativo nei due casi: se le IT sono sempre state uno strumento atto al miglioramento dell'efficienza e dell'efficacia interne, le ICT – al contrario – danno il meglio di sé nel supporto all'interazione inter-organizzativa, favorendo e sostenendo lo sviluppo dell'*e-business* (dove per *e-business* si intende l'integrazione a monte e a valle dell'impresa, cioè la collaborazione in un *network* con valenze strategiche, e non il mero "commercio elettronico").

È poi importante ricordare che, anche per i motivi poc'anzi illustrati, le ICT possono avere un impatto positivo sulle performance della PMI italiana solo a patto che esista un adeguato *fit* tra la strategia di business e quella impiegata per la definizione degli investimenti in tecnologie.
A questo scopo è essenziale, da parte del vertice strategico, una profonda comprensione delle potenzialità offerte dalle ICT, al fine di sviluppare strategie coerenti. Tuttavia, si rileva che la PMI quasi mai dispone degli *skill* interni sufficienti per sviluppare una *vision* adeguata.
Questa situazione è rispecchiata nel forte ritardo nell'adozione di ICT registrato dalla PMI italiana: a fronte di numerosità e di forte impatto sul PIL, essa è infatti quella che investe di meno, e gli investimenti rappresentano in genere solo migliorie incrementali.

L'analisi – teorica ed empirica – delle cause che stanno alla base del ritardo tecnologico della PMI italiana pone in evidenza l'esistenza di due macro-categorie di problematiche, che si collocano l'una lato domanda (cioè PMI) e l'altra lato offerta (cioè fornitori di ICT operanti con la PMI), riassunte qui di seguito.

**Problematiche lato domanda**:

- scarsa comprensione delle <u>opportunità</u> connesse alle tecnologie adeguate alle necessità delle PMI: questo impedisce alle PMI di superare i *gap* di *performance* o di esplorare nuove opportunità;
- scarsa comprensione di <u>come implementare</u> adeguatamente le tecnologie: questo comporta una scarsa capacità di perseguire efficienza, efficacia e innovazione;
- scarso <u>presidio manageriale</u>: il numero delle PMI che dispongono di una direzione (o di responsabile) ICT è esiguo;

- carenza appropriate competenze nelle risorse umane (*skill shortage*): tipicamente una PMI dispone di uno staff ridottissimo (o non ne dispone affatto!) dedicato alle ICT;
- difficoltà nel valutare il ROI (ritorno sull'investimento) e ridotta capacità di sviluppare una vision strategica per gli investimenti in ICT, derivante dalla concomitanza di tutti o parte dei punti precedenti;
- costo della tecnologia

Inoltre, si rileva che i problemi elencati qui sopra possono essere aggravati in molti casi dalla relazione intercorrente tra:

- Età e background del decisore di acquisto e "strategicità" dell'investimento;
- Posizione competitiva dell'impresa e propensione all'innovazione in ICT;
- Settore, classe dimensionale, area geografica e livello adozione ICT.

**Problematiche lato offerta:**

- *lock-in*: tentativo, da parte del fornitore, di legare a sé l'impresa cliente (ad es. attraverso applicativi sviluppati ad hoc) anche a scapito della reale efficacia delle soluzioni implementate;
- prodotti inadeguati alle esigenze: spesso i prodotti proposti alle PMI non sono adeguati alle loro reali esigenze (ad esempio possono essere sovradimensionati, o non in grado di offrire funzionalità effettivamente efficaci per le esigenze della specifica impresa);
- incapacità di comprendere i clienti: spesso i fornitori di ICT non sono in grado di comprendere a fondo le problematiche, le esienze e le specificità della singola PMI cliente, e finiscono con il sottovalutare le componenti umana, strategica o organizzativa nella progettazione di soluzioni informatiche, che finiscono così con il risultare inefficaci;
- Immaturità: incapacità, da parte dei fornitori di ICT per la PMI, di proporsi come partner tecnologico a tutto tondo, capace di offrire non solo tecnologia, ma anche e soprattutto consulenza.

**Figure fondamentali per la PMI che vuole adottare ICT**

Infine è importante ricordare che, affinché l'innovazione tecnologica abbia luogo e sia costruita su solide basi (presupposto per un impatto positivo sulle performance aziendali), è necessario che la PMI ponga estrema attenzione nella selezione di due attori in grado di giocare un riuolo chiave:

- il "partner tecnologico a tutto tondo": dovrebbe essere possibilmente un *system integrator*. Il suo ruolo è quello di fornire un supporto consulenziale alla PMI nella scelta e progettazione di investimenti in ICT;
- il "pivot dell'innovazione": può essere interno o esterno, ma deve guidare la vision degli investimenti in ICT nella PMI.

**PARTE 2**

**PROBLEMATICHE ECONOMICO-ORGANIZZATIVE
LEGATE AL PARADIGMA OPEN SOURCE**

## 6. Il problema della sostenibilità economica del modello *open-source*, e dello stimolo all'innovazione da esso offerto.

Non è possibile evitare almeno di accennare alle dibattute problematiche connesse alla sostenibilità economica e organizzativa del modello *open-source* nel medio-lungo periodo, e di quanto esso sia effettivamente in grado di sostenere e stimolare l'innovatività. Scopo di questo capitolo è, perciò, non tanto quello di indicare delle velleitarie "soluzioni ottime", quanto piuttosto quello di offrire degli spunti di riflessione da utilizzare per una costruttiva analisi contestuale da parte di chi si accinge ad intraprendere la strada dello sviluppo di un *business* basato sul *Free/OpenSource Software* (F/OSS).

Il fenomeno del *Free/OpenSource Software* genera dibattiti molto accesi tra suoi fautori e suoi detrattori, il più delle volte con contrapposizioni che sono di fatto "ideologiche" e sostenute o da tesi piuttosto generiche che tendono a fornire una spiegazione per la "totalità" del fenomeno F/OSS, senza però indagarne le peculiarità, o da tesi economiche che non considerano le speciali caratteristiche del bene "software" e del relativo mercato.

Alcuni fenomeni che meritano attenzione e che stanno – in tutto o in parte – alla base del fenomeno F/OSS sono:

– la *crisi dei modelli di produzione tradizionali*, che implicavano una posizione di dominio da parte del produttore sul "consumatore" (l'utente nel caso del software) finale;
– il sorgere di una *comunità di sviluppatori* che esce dai tradizionali schemi di produzione industriale e, a volte, anche dalle regole economiche del mercato;
– delle *reazioni estreme*, anche se giustificabili, alla presenza di *posizioni fortemente monopolistiche* in settori di largo consumo, strategici per il mercato ma anche per la società moderna, come quello dei sistemi operativi e degli strumenti di *office automation*.

Uno degli argomenti maggiormente dibattuti che differenzia l'approccio proprietario e quello F/OSS riguarda la *sostenibilità economica* di quest'ultimo nel medio–lungo termine (che è poi quel periodo durante il quale vengono generati ritorni sugli investimenti, e che nei mercati più "tradizionali" alimenta l'occupazione nel settore, stimola l'innovazione e gli ulteriori investimenti, ecc.). In questo ambito vi sono delle posizioni fortemente contrapposte: da una parte i produttori di software che vedono, ad esempio, nella Gnu Public Licence un "mostro" capace di distruggere il mercato e soffocare l'innovazione, e che paventano nelle questioni di proprietà intellettuale un forte freno all'accettazione di prodotti F/OSS da parte delle realtà imprenditoriali (cioè dalla fascia di mercato "*Enterprise*"), dall'altra la posizione altrettanto "integralista" del mondo F/OSS che affonda la sue radici in motivazioni spesso maggiormente legate alla *self-promotion*, al prestigio personale, alla voglia di sfida, al mutuo soccorso, ecc. (si assiste cioè al passaggio dal paradigma "*good for money*" a "*gift for reputation*", come evidenziato dal Boston Consulting Group) che alla conoscenza delle leggi che governano il mercato (non si dimentichi che quello del software **è** un mercato), le reazioni dei consumatori, ecc.

### 6.1 Caratteristiche e struttura del mercato del software.

Dato che con difficoltà possiamo pensare di trasformare il software a una sorta di "bene pubblico" (a meno di non patire poi le conseguenze derivanti dall'azzeramento della relativa industria), è necessario definire dei modelli di sostenibilità validi per garantire redditività anche a chi adotti approcci F/OSS, dato che quello del software è uno dei maggior *driver* dell'economia mondiale. Per perseguire questo obiettivo è necessario innanzitutto concentrarsi su quelle che sono le caratteristiche fondamentali e la struttura del mercato software (così come analizzate, ad es., da MICE, 2003), riportati per sommi capi in Tab. 11 e 12.

| Tabella 11 - Caratteristiche del mercato del software (tradizionale) ||
|---|---|
| **Lato Offerta** | **Lato domanda** |
| a) *costi di sviluppo molto alti*, con elevati costi di produzione sommersi ("sunk costs"), che si verificano durante lo sviluppo e prima della commercializzazione; | a) *Obsolescenza*: anche se è un bene immateriale senza decadenza fisica o scadenza, il software è soggetto ad obsolescenza molto rapida; |
| b) *costi marginali di produzione estremamente bassi*: superato il costo iniziale elevato di R&D, dato il bassissimo costo di replicazione, il costo marginale del software è estremamente basso: di conseguenza anche i costi medi assumono un andamento fortemente decrescente al crescere della produzione, rispetto ad un bene non digitale; | b) *Effetto di rete*: come già riportato per quanto riguarda l'offerta; |
| c) *economie di scopo*: i singoli elementi che compongono il software possono essere utilizzati in prodotti anche estremamente diversi fra loro; | c) *Effetto Lock-in*: l'utente che ha supportato dei costi per utilizzare il software sarà molto contrario a cambiarlo, a causa degli ulteriori costi di apprendimento. L'effetto lock-in può essere considerato una barriera all'entrata per i concorrenti; |
| d) *effetto rete*: si basa pertanto sul famoso postulato di Metcalfe, secondo il quale il valore di un | d) *non concorrenza nel consumo*: un software può essere fisicamente installato allo stesso modo su |
| network cresce esponenzialmente con il numero dei nodi connessi. Risulta fondamentale creare e stabilire un certo standard che permetta la comunicazione fra il più alto numero possibile di fruitori; | infinite postazioni e usato conseguentemente da diversi utilizzatori creando una non concorrenza nel consumo. |
| e) *intangibilità del bene*, a causa della quale è difficile una corretta valutazione economica del software, soprattutto ex-ante: il problema è ben noto nell'ambito del Software Engineering; | |
| f) *internazionalizzazione:* data la non fisicità del software è evidente la sua grande facilità di distribuzione che interessa il mercato internazionale e globale; | |
| g) *non consumabilità*: l'offerta nel tempo non si esaurisce poiché nel mercato vi è l'immissione costante di nuovi prodotti e versioni. | |

| **Tabella 12 - Struttura del mercato del software** | |
|---|---|
| *competizione*: bassi costi marginali e alti costi fissi di R&D creano nel mercato una forza di rinnovamento continua che rende possibile la vendita del numero maggiore di copie di un software nel minor tempo possibile, al fine di finanziare in maniera adeguata i costi di sviluppo. L'innovazione percorre la strada della differenziazione (sia di prodotto che di mercato) soprattutto nel caso di mercati ormai saturi. Vi è inoltre un vantaggio nell'attuare operazioni di concentrazione per cercare di ridurre l'ammontare dei costi fissi e di ampliare il mercato di riferimento. | |
| *barriere all'entrata*: un'impresa che voglia entrare in questo mercato incorre in costi di R&D molto elevati, ma allo stesso tempo deve sostenere dei costi di entrata estremamente bassi se paragonati a quelli sostenuti dalle imprese che vendono beni "fisici" (e che quindi necessitano della creazione di impianti produttivi). Per questo motivo, entrate e uscite sono continue, e vi è la possibilità che anche piccole realtà imprenditoriali riescano a ricavarsi uno spazio significativo (come nei casi storici di Netscape™, Napster™, lo stesso MS-DOS™). | |
| *standard e licenze*: la creazione di standard comuni e applicati su larga scala crea vantaggi enormi sia per l'usabilità del software, sia in termini di economie di scala e diminuzione dei costi per l'interconnessione di sistemi con standard differenti. | |
| *Marketing – caso proprietario:* curato dall'impresa che sviluppa il programma<br>– prodotto: enfasi sul Time-to-Market (rientro dei costi R&D)<br>– prezzo: basato sulle strategie di marketing<br>– distribuzione: tradizionale<br>– pubblicità: investimenti "tradizionali" | *Marketing – caso F/OSS:* non è individuabile un soggetto preciso che segue la funzione di marketing (eccetto F/OSS inglobato in aziende che hanno un brand importante)<br>– prodotto: enfasi sulla qualità<br>– prezzo: difficilmente utilizzabile come leva competitiva<br>– distribuzione: soggetta a meccanismi particolari<br>– pubblicità: meccanismi particolari (non è possibile coprire i costi promozionali con il prezzo), es. sfruttamento di un marchio affermato, fondazioni, passa parola, ecc. |

È interessante, in particolare, focalizzarsi sulle peculiarità legate alla determinazione del *prezzo*, elemento che – dal punto di vista della struttura del mercato - maggiormente distingue il paradigma F/OSS da quello proprietario. Come ben noto, in un'economia di mercato la determinazione del prezzo avviene attraverso meccanismi di coordinamento che hanno luogo nell'incontro di domanda e offerta. Il prezzo, però, assolve anche a tutta un'altra serie di funzioni, tra le quali una delle più importanti è quella di fungere da meccanismo sulla base del quale si ottiene la corretta allocazione delle risorse (le variazioni di prezzo tendono infatti a indirizzare le risorse per rispondere alle necessità). È facile intuire che, nel modello F/OSS, la mancanza di un prezzo, unita alla mancanza di valutazione esplicita dei costi-opportunità (gli sviluppatori F/OSS rinunciano a parte del proprio tempo libero per occuparsi dello sviluppo di software) può portare a un'allocazione non soddisfacente delle risorse e al conseguente disallineamento tra offerta e domanda (in altre parole i criteri e le priorità di sviluppo rischiano di non corrispondere a quelle che sono le reali esigenze e urgenze della potenziale domanda). In conclusione, perciò, il modello F/OSS "puro" per quanto riguarda l'allocazione delle risorse e la valorizzazione dal punto di vista del prezzo, non è in generale né efficiente (è impossibile effettuare una comparazione diretta tra input e output) né efficace, in quanto non vi sono meccanismi per misurare direttamente il grado di ottimizzazione dell'allocazione delle risorse. Questi limiti vengono però superati quando il modello F/OSS si combina con logiche che comportano l'adozione di un prezzo: questo implica però anche la necessità di rientrare in logiche di mercato "tradizionale"…

**6.2 Modello F/OSS: gestione degli *asset* aziendali**

Il turbolento ed altamente competitivo mercato del software tende ad articolarsi secondo due principali approcci per lo sviluppo del prodotto: quello "a cattedrale", fondato sulla centralità della pianificazione ed esecuzione in un'organizzazione *top-down*, tipica del software proprietario, e quello "a bazar", fondato invece sulla decentralizzazione, basata su di un'organizzazione a rete, che crea una comunità virtuale di sviluppatori, come nel caso del F/OSS. Esistono anche tipologie ibride (più vicine a quanto poi avviene nella realtà), come il *cooking-pot* (Ghosh, 1998).
Da analisi recentemente svolte dal Boston Consulting Group è possibile derivare quali siano le motivazioni e le gratificazioni che spingono gli sviluppatori a collaborare a progetti di F/OSS (cfr. Tab. 13).

| Tabella 13 - Motivazioni e gratificazioni per gli sviluppatori F/OSS ||
|---|---|
| **Motivazioni** | **Forme di riconoscimento** |
| Attività non retribuita, svolta nel tempo libero, con obiettivi strettamente personali:<br><br>− crescita delle conoscenze<br>− miglioramento della posizione professionale<br>− posizioni ideologiche sulla libertà di accesso al codice | − lo sviluppatore contribuisce in parte a creare/migliorare F/OSS, quindi un costo minimo rispetto al miglioramento complessivo acquisito attraverso la somma dei lavori di ogni componente della comunità;<br>− lo sviluppatore rimane sempre aggiornato con percorsi di training continuo;<br>− lo sviluppatore riesce a mostrare le sue abilità molto più velocemente rispetto alle vie tradizionali ("signaling effect"). |

È importante rilevare che gli elementi di soddisfazione e gratificazione potrebbero (dovrebbero) essere tenuti in grande considerazione dalle *software house* (comprese quelle che si occupano di software commerciale) quali strumenti per il contenimento del *turnover* e la qualificazione del personale.

## 6.3 Lati negativi dello sviluppo F/OSS

| Tabella 14 - Problemi legati al modello di sviluppo F/OSS |
|---|
| *Mancato incontro di domanda e offerta*: <br> molti software F/OSS vengono sviluppati al fine di risolvere i problemi degli sviluppatori, che però possono non corrispondere alle reali esigenze del mercato |
| *Virtuosismo*: <br> a volte vengono sviluppati software F/OSS estremamente complicati che hanno la fondamentale finalità di dimostrare la bravura dello sviluppatore (e quindi accrescere la sua reputazione) |
| *Happy engineering*: <br> il processo di sviluppo non considera come primarie caratteristiche fondamentali quali: l'usabilità, l'operatività, il valore aggiunto e i possibili costi aggiuntivi |
| *Mancanza di valore aggiunto per l'utilizzatore*: <br> il fine ultimo di un prodotto software dovrebbe essere quello di creare valore aggiunto per chi decide di utilizzarlo. La mancanza di prezzo rende in generale più difficoltoso il processo di "aggiustamento" tra domanda e offerta (in effetti è necessario considerare che non sempre la soluzione più sofisticata è la "migliore": tanto maggiore è il valore percepito dall'utente, e tanto maggiore sarà il profitto). Raramente nel F/OSS si tiene conto di un utilizzo "di massa" del software, e, in questo modo, raramente i bisogni della maggioranza dei possibili utenti vengono soddisfatti. |

## 6.4 Modello F/OSS e stimolo all'innovazione.

Un altro aspetto fortemente dibattuto riguardo il modello F/OSS è quello legato alla sua attitudine a stimolare e generare innovazione. Se da una parte alcuni eminenti autori sottolineano la capacità del paradigma OS di generare spinte innovative *user-driven* (cfr. ad es. Von Hippel, 2001, 2002), dall'altra non mancano le critiche, basate su argomentazioni che non sono da considerarsi del tutto prive di fondamento. Dato che, nel campo del F/OSS i benefici sono generalmente più noti delle critiche, ci sembra interessante cominciare con l'elencare nella seguente tabella almeno le principali tra di esse, che rappresentano altrettante aree di attenzione per l'impresa che decidesse di investire in F/OSS.

| Tabella 15 - Limiti alla capacità di stimolare l'innovazione da parte del modello F/OSS |
|---|
| Benché in alcuni F/OSS la spinta innovativa sia reale, lo sviluppo di molti altri è in realtà finanziato da grandi *player* che sperano di ottenere abbondanti rientri nel lungo periodo derivanti dall'affossamento del monopolista. |
| Se consideriamo i principali prodotti OS diffusi sul mercato (Linux, Open Office, The Gimp ecc.), possiamo notare che in realtà si tratta non di applicazioni innovative, bensì di alternative *free* a prodotti commerciali ampiamente consolidati (che probabilmente hanno costi eccessivi per il grande pubblico). |
| Evans e Reddy (2002) affermano: "*Clearly, much innovation in commercial software has occurred over those 25 years. Just as clearly, much (but certainly not all) of the focus of GPL software over the past two decades has been on creating "free" versions of proprietary software*". Essere la versione gratuita e/o a codice aperto di un prodotto affermato è chiaramente un elemento di marketing molto forte … |
| Il software F/OSS stimola i comportamenti opportunistici (*free-riding*) da parte di chi beneficia |
| gratuitamente di beni che possono essere considerati alla stregua di beni pubblici. |
| Scarso interesse, da parte degli sviluppatori *open source*, ad assumere atteggiamenti di *first-mover* (cosa invece perseguita dalle aziende, che si trovano nella necessità di acquisire e difendere vantaggio competitivo) |
| Caduta di interesse nel tempo per lo sviluppo e l'innovazione continua di uno specifico software, da parte della comunità di sviluppatori. |
| Scarsa convergenza tra obiettivi di sviluppatori e utenti finali. |
| Il libero accesso al codice sorgente favorisce i fenomeni di imitazione, e porta al proliferare di prodotti molto simili dotati sostanzialmente di prestazioni allineate (caso esemplare è quello dei Content Managemetn System – CMS, di cui esistono innumerevoli versioni, spesso poco differenziate tra loro). |
| Scarsa capacità di creare modelli di business realmente innovativi (nella maggior parte dei casi si assiste al riutilizzo di modelli consolidati, basati ad es. sulla vendita di servizi aggiuntivi, sulla consulenza, sull'inserimento di messaggi promozionali, ecc.) |

### 6.5 I modelli di business basati su F/OSS

La mancanza di ritorno economico connaturata nel F/OSS può essere bilanciata, come precedentemente esplicato, attraverso la creazione di attività o settori complementari: vendita di servizi, software, hardware aggiuntivi (ovviamente l'attività diviene più redditizia, quanti più servizi aggiuntivi si è in grado di supportare). In particolare la stessa comunità F/OSS ha identificato un serie di modelli (utilizzati *as-is* oppure in forme ibridate) che dovrebbero garantire la sostenibilità economica, che riassumiamo in Tab. 16; la tabella riporta anche un'ulteriore serie di modelli di business (in parte sovrapponibili con i precedenti) elaborata in MICE, 2003.

| **Tabella 16 - Modelli di business F/OSS** |
|---|
| **Modelli di Business proposti dalla comunità F/OSS** |
| *Support seller:*<br>il software viene distribuito in forma F/OSS, ma si vendono elementi quali la distribuzione, il brand, i servizi post-vendita… |
| *Loss Leader:*<br>il produttore usa F/OSS per entrare nel mercato e avere visibilità; poi ci si espande applicando prodotti proprietari al software originale. |
| *Widget frosting:*<br>il F/OSS è usato dai produttori hardware per facilitare il debugging complesso, come i driver per le periferiche, con lo scopo di fornire un package (hardware + software) migliore ad un costo decisamente inferiore. |
| *Accessorizing:*<br>vendere accessori che variano da gadget e merchandising a interi prodotti hardware, compatibili alla idee della comunità open-source. |
| *Software franchising:*<br>poco utilizzato perché eccessivamente prono a comportamenti opportunistici. |
| **Modelli di business proposti in MICE, 2003** |
| *Vendita di servizi addizionali:*<br>aggiungere al software servizi per adattare il prodotto alle varie esigenze utente. Non è da trascurare come tale sviluppo vada a ricadere sull'intera comunità, e se da un lato questo fatto porta alla crescita del prodotto, dall'altro espone le imprese già presenti nel mercato all'entrata di nuove aziende che possono facilmente collocarsi senza barriere di ingresso, usufruendo degli investimenti effettuati dai soggetti già presenti; |

> *Vendita di software addizionali:*
> il F/OSS è la base sulla quale, con le apposite licenze, sviluppare e vendere software commerciali e relativi servizi. Questo approccio si scontra con limitazioni non tanto economiche, ma di principio: infatti chi aderisce al movimento F/OSS, crede nella libera consultazione del codice sorgente e nella distribuzione gratuita; difficilmente, perciò, sarà disposto a sviluppare applicazioni commerciali.
> Inoltre è da segnalare che non sempre è facile ottenere profitti soddisfacenti basandosi solo su questo approccio, a meno di non detenere una fetta consistente del mercato: tali imprese, infatti, possono investire in modo redditizio, soprattutto nel caso adottino una piattaforma - rigorosamente open-source - sulla quale inserire software addizionale. Così facendo non si sostengono i costi necessari a disporre di una piattaforma stabile, e i margini derivanti dalla vendita combinata con il software addizionale sono più sostanziosi.

> *Vendita di hardware correlato a F/OSS:*
> questo modello può essere adottato sostanzialmente solo da imprese che promuovono la vendita di F/OSS a corredo del proprio hardware (la strada tracciata da IBM™ su larga scala - e in un settore ad elevata criticità come quello dei server ne è un lampante esempio). Si può ottenere software completamente integrato, uniforme e compatibile con l'hardware. Tra l'altro, l'utilizzatore non si trova costretto a sostenere il costo di acquisto per le licenze del software commerciale, disponendo così di un surplus di denaro che può eventualmente utilizzare per acquistare un hardware più costoso (e perciò potenzialmente più redditizio per il fornitore).

È necessario evidenziare che, da una serie di ricerche condotte recentemente, è emerso che, data la centralità delle motivazioni ideologiche degli sviluppatori nell'aderire a comunità *open-source*, è difficile reperire soggetti che si prestino a rendere commerciali parti/personalizzazioni del software. È quindi necessario, per le aziende che volessero investire in soluzioni F/OSS, dotarsi (con i costi che ne derivano) di *un gruppo di sviluppatori permanenti*, similmente a quanto avviene nel caso del software proprietario.

Per finire, si elencano alcune aree di attenzione che sono segnalate dalle più recenti ricerche per chi intenda adottare un modello di business che preveda l'uso di F/OSS:

- è importante distinguere le applicazioni per categorie (es.: sistema operativo, *office automation*, *utility* ecc.), e vagliare attentamente quali tra esse meglio si adattino ad essere implementate utilizzando F/OSS (se è evidente che il sistema operativo può essere F/OSS, non è detto che ciò sia sempre la scelta migliore, ad es., per un'applicazione di *e-commerce* fortemente personalizzata);

- è bene contestualizzare il prezzo, diversificandolo sulla base di considerazioni legate all'utente finale, alla dislocazione geografica, al mercato di riferimento, alla situazione sociale ecc., al fine di inglobare le giuste istanze sociali che costituiscono un elemento di forte motivazione per il F/OSS;

- è bene ricordare che, in generale, un software che deve accedere ad Internet crea una condizione nella quale il grado di "apertura" del codice dovrebbe essere maggiore rispetto alla stessa categoria di software senza apertura verso la rete, soprattutto al fine di garantire sicurezza e *privacy*.

## 7. Open source e organizzazione: come migliorare il processo di produzione del software

Lo sviluppo di sistemi software di grandi dimensioni è un'attività complessa, che implica rischi sia manageriali che tecnici. Nel paradigma tradizionale di sviluppo sono stati profusi grandi sforzi (tecniche di *project management*, modularizzazione dei progetti, ecc.) nel tentativo di ottenere software "di qualità", a volte con la conseguenza "imprevista" di allungare i tempi di sviluppo. Per mantenere la competitività nell'era di Internet, tuttavia, la qualità del software deve necessariamente essere accompagnata da un pronta reattività e da tempi di sviluppo ridottissimi.

Una tra le possibili soluzioni ai problemi organizzativi che affliggono l'industria del software è l'adozione di modelli di organizzazione del lavoro basati sul modello delle *community* F/OSS: si tratta – come propongono Sharma, Sugumaran e Rajagopalan - di creare *community* di sviluppatori "ibride" in grado di ricreare parte delle condizioni proprie di una *community* di sviluppatori F/OSS nel più rigido paradigma di un gruppo di lavoro per lo sviluppo di software in contesto aziendale. È tuttavia importante sottolineare che questa soluzione non può assolutamente essere applicata indiscriminatamente a qualsiasi contesto (tra l'altro è cosa nota che il successo dei progetti F/OSS si è registrato prevalentemente nelle applicazioni orizzontali - **infrastrutture software** *general purpose* -).

Per comprendere appieno quali siano gli aspetti che un'azienda intenzionata a basare il suo business sull'adozione di soluzioni F/OSS deve considerare con attenzione, è importante considerare quale sia l'impatto organizzativo derivante da un'organizzazione del lavoro basata su principi che si fondano nel paradigma di *community* e di "*gift for reputation*". È abbastanza intuitivo pensare che possano sorgere dei problemi di coordinamento, ad esempio, tra la *community* degli sviluppatori "volontari" e gli sviluppatori che invece, come abbiamo visto, è necessario assumere, dato che questi ultimi saranno verosimilmente sottoposti a vincoli differenti (es. scadenze di progetto, necessità di soddisfare specifiche esigenze dei clienti, *time to market* stringenti, ecc.) e orientati al raggiungimento di obiettivi (es. rispetto di parametri di produttività, incentivi economici, ecc.) non necessariamente coincidenti con quelli della *community* (in genere finalizzati alla costruzione della reputazione).

Diversi ricercatori si sono occupati di questi aspetti. Nel seguito riassumiamo alcune tra le loro osservazioni maggiormente rilevanti.

**7.1 Il cambiamento paradigmatico introdotto dal F/OSS**

Il "fenomeno" F/OSS sta portano a un cambiamento paradigmatico nell'industria del software, che si sta vieppiù trasformando, passando da industria di produzione a industria di servizio. Ciononostante le organizzazioni *for-profit* stanno sperimentando delle concrete difficoltà nel costruire modelli di business adeguati a sfruttare il paradigma F/OSS. Inoltre non esiste ancora un corpus di ricerca davvero approfondito e completo su queste problematiche.

La "crisi" dell'industria software – sempre alla ricerca di nuovi modi per creare nuovi prodotti di qualità elevata – ha trovato nell'*open source* le potenzialità offerte da una tecnologia *innovativa*, *flessibile*, *veloce* e a *costo contenuto*. E, di fatto, il F/OSS ha aiutato le imprese a stabilire *standard* industriali, a migliorare il grado di *penetrazione del mercato* e il *vantaggio competitivo*, e a *fidelizzare* i clienti e gli sviluppatori, che si sentono *motivati* dal senso di "proprietà" sul software derivante dal poter accedere al codice sorgente. Inoltre, il F/OSS – senza alcuna campagna marketing – si è guadagnato rilevanti posizioni di mercato in differenti ambiti (e.g. Apache).

### 7.2 F/OSS, struttura organizzativa, processi e cultura organizzativa

Come ben noto nelle scienze organizzative, ogni organizzazione è il compendio specifico di una *struttura*, una *cultura* e un insieme di *processi*. Nelle tabelle seguenti (Tab. 17) sono perciò riassunte e confrontate le principali caratteristiche che contraddistinguono, da punto di vista di questi tre aspetti, le organizzazioni "tradizionali" e le *community* F/OSS; le tabelle contengono altresì le caratteristiche specifiche che dovrebbero essere presenti nelle comunità di sviluppatori "ibride" (cioè calate in una realtà aziendale) cui si è fatto cenno nell'introduzione a questo paragrafo.

**Tabella 17 -** Confronto tra le strutture organizzative tradizionali e le community F/OSS

| | Forme organizzative tradizionali | | | | |
|---|---|---|---|---|---|
| | **FUNZIONALE** | **DIVISIONALE** | **MATRICE** | **OSS COMMUNITY** | **HYBRID-OSS COMMUNITY** |
| **Divisione del lavoro** (assegnamento dei task) | Per input | Per output | Per input ed output | Per scelta e conoscenza | Assegnamento volontario, basato su skill, competenze, conoscenza |
| **Meccanismi di coordinamento** | Supervisione gerarchica, piani e procedure | Direttore generale della divisione, staff | Doppio reporting | Gestione dell'appartenenza, regole e istituzioni, monitoraggio, controllo e sanzioni | Monitoraggio tra pari, gestione dell' appartenenza basata sulla reputazione, su regole e istituzioni |
| **Potere decisionale** | Fortemente centralizzato | Separazione tra strategia ed esecuzione | Condiviso | Altamente democratico e decentralizzato | Basato sul consenso, con processi decisionali decentralizzati |
| **Confini organizzativi** | Centro /periferia | Mercati interno/esterno | Interfacce multiple | Porosi e in evoluzione continua | Permeabili, governati da regole organizzative e da istituzioni |
| **Importanza strutture informali** | Bassa | Contenuta | Considerevole | Alta (non esiste struttura formale, ma al più un gruppo core) | Reti informali |
| **Struttura** | Inter- | Corporate- | Lungo le | Coalizioni variabili | Questioni rilevanti |

| politica | funzionale | divisioni e inter-divisionale | dimensioni della matrice | | per i membri dell'organizzazione |
|---|---|---|---|---|---|
| **Legittimazione dell'autorità** | Expertise derivante dalla posizione e dalla funzione | Risorse e responsabilità del general management | Negoziazione di skill e risorse | Reputazione (meritocrazia) | Reputazione, basata su di un periodo sufficientemente ampio |

**Tabella 18 -** Confronto tra i processi (organizzazioni tradizionali vs. community F/OSS)

| | **ORGANIZZAZIONE TRADIZIONALE** | **OSS COMMUNITY** | **HYBRID-OSS COMMUNITY** |
|---|---|---|---|
| **Processo di governance** | Rinforzo alla governance | Self-governance (basata su team che si occupano di moduli e gruppo core che coordina il tutto) | Potenziamento |
| Meccanismi di governance: | | | |
| *Gestione appartenenza* | - rinforzo management<br>- statica/solida | - basata sulla community<br>- basata sulla verifica della qualità del membro<br>- fluida, ma stabile<br>- identità professionale | Selezione di persone qualificate, assegnazione compiti |
| *Regole e istituzioni* | Il management crea e modifica le regole | I membri della community creano e modificano le regole (meccanismi di voto con verifiche tra pari) | Facilitare la creazione di regole e norme, garantire autonomia |
| *Monitoraggio e sanzionamento* | - monitoraggio performance e comportamenti (tenuti confidenziali)<br>- sanzioni economiche, retrocessione, layoff, ecc. | - monitoraggio performance e comportamenti (pubblico)<br>- sanzioni sociali: flaming, spamming, espulsione, ecc.<br>- creazione e mantenimento reputazione come fattore primario di motivazione | - supporto open, monitoraggio tra pari<br>- garantire autorità per rinforzare sanzioni e premi |
| *Reputazione* | Nessuna enfasi sulla costruzione di reputazione | Perdita/aumento reputazione come fattore motivante | Riconoscere qualità del lavoro e promuovere la reputazione |
| **Processo di sviluppo** | | | |
| *Fasi del processo* | - survey<br>- studi<br>- definizione<br>- configurazione<br>- approvvigionamento<br>- design<br>- costruzione<br>- consegna | - riconoscimento problemi<br>- ricerca volontari<br>- identificazione soluzioni<br>- sviluppo e testing codice<br>- review modifiche codice<br>- documentazione e rilascio codice<br>- gestione release | - generazione idee, identificazione opportunità<br>- assegnazione compiti a persone con appropriato background<br>- abbandono metodologie sviluppo tradizionali<br>- supporto allo sviluppo e al testing di codice autonomi<br>- creazione di struttura di supporto al coordinamento<br>- allocazione risorse per il |

|  |  |  | rilascio e il supporto del prodotto |
|---|---|---|---|

**Tabella 19 -** Confronto tra le culture (organizzazioni tradizionali vs. community F/OSS)

|  | ORGANIZZAZIONE TRADIZIONALE | OSS COMMUNITY | HYBRID-OSS COMMUNITY |
|---|---|---|---|
| **Artefatti:** | | | |
| *Comunicazione* | Face-to-face | Mediata dal computer | È incoraggiata la comunicazione elettronica tra gruppi con diverse localizzazioni |
| *Localizzazione* | Multiple | Multiple, globali, multiculturali | |
| **Valori:** | | | |
| *Rischio* | Manager/proprietario | Condiviso nella community | |
| *Proprietà* | Manager/proprietario | Condivisa dalla community | |
| *Ricompense* | Favoriscono la proprietà | Basate sul merito e la condivisione | Riconoscimento dell'expertise |
| *Motivazione* | Soprattutto economica | Altruismo, reputazione ideologia | Set di motivazioni "allargato" |
| *Informazione* | Condivisa sulla base delle necessità | Condivisa apertamente | Aperta condivisione tra i gruppi di lavoro |
| *Decision making* | Autocratico | Generalmente democratico (votazioni) | Auto-gestione dei lavoratori della conoscenza |
| *Controllo* | Mantenuto da decisori autonomi | Regole di appartenenza, licenze, procedure di voto | |
| *Strutture di lavoro* | Rigide | Flessibili | |
| **Assunzioni core:** | | | |
| *Trust* | Non basata sul trust | Basata sul trust reciproco | È necessario incoraggiare lo sviluppo del trust reciproco tra gli sviluppatori core |
| *Lealtà* | Non basata sulla lealtà | Lealtà condivisa | |

### 7.3 Come le imprese possono sfruttare l'organizzazione "stile OS"

Il modello organizzativo delle community ibride è ormai adottato con successo da diversi grandi *player* (Sun Microsystems, IBM, Intel, ecc.) e dovrebbe essere considerato tra le opportunità di riorganizzazione per tutte quelle imprese – indipendentemente dalla dimensioni - che si affidano a un processo di sviluppo software ancora basato su paradigmi tradizionali simili a quello manifatturiero. Tali aziende dovrebbero chiedersi in quali aree o progetti potrebbero essere applicato con successo il modello ibrido, al fine di cogliere alcuni tra i seguenti vantaggi:

- riduzione del tempo di sviluppo e del *time-to-market*
- miglioramento della qualità
- riduzione dei costi
- fidelizzatine del personale sviluppatore

- sviluppo professionale del personale sviluppatore senza *overhead* onerosi (es. per corsi di aggiornamento).

È tuttavia necessario tener presente che la transizione verso questo modello organizzativo innovativo richiede da parte del management e del personale:

- la comprensione profonda della filosofia F/OSS
- lo sviluppo della fiducia reciproca tra management e sviluppatori
- la percezione da parte degli sviluppatori, di essere coinvolti in progetti innovativi e impegnativi
- una forte motivazione da parte degli sviluppatori nel partecipare ai progetti

**7.4 La gestione dei conflitti strutturali nelle community "ibride"**

Alcuni autori, tra i quali ad esempio van Wendel de Joode, si sono occupati delle problematiche derivanti dalla necessità di gestire i conflitti strutturali che tendono a sorgere nell'ambito di community di sviluppatori "miste" (cioè che includono anche soggetti appartenenti ad ambiti business).
Si tratta di un problema non banale, e spesso sottovalutato, perchè la tendenza generale è quelle di sottolineare soprattutto i vantaggi derivanti dal paradigma organizzativo F/OSS, particolarmente in termini di creazione e condivisione di conoscenza e innovazione. Vi sono addirittura esempi di organizzazioni che erigono di proposito le community F/OSS a modello per l'innovazione (è il caso, ad esempio, della Philips).
Tuttavia è necessario avere ben presente che le community di sviluppatori OS sono composte da professionisti - appassionati di sviluppo e desiderosi di apprendere – che operano al di fuori di controlli gerarchici stringenti e senza dover rispettare deadline precise e predefinite. Le aziende, però, desiderano avere il controllo dei loro processi e dei risultati, ed esigono il rispetto delle sepcifiche e delle tempistiche. La nascita di conflitti è perciò difficilmente evitabile.

Un progetto finanziato dall'olandese Netherlands Organisation for Scientific Research (NWO), ha recentemente indagato proprio queste problematiche, analizzando tra gli altri il caso del *Content Managment System* (CMS) MMbase. Questo caso di studio è per molti versi paradigmatico della tipologia di conflitti che possono sorgere attorno a software OS che raccoglie l'interesse del "mercato". È quindi interessante capire come i conflitti strutturali sono sorti e sono stati gestiti.
MMbase è stato inizialmente sviluppato dall'emittente pubblica olandese, che ha poi deciso di distribuirlo pubblicamente, sia per motivi di coerenza etica (lo sviluppo di MMbase era stato finanziato con denaro pubblico), che per cercare un supporto esterno nel mantenimento dell'applicazione.
Il nuovo paradigma è stato all'origine di una serie di cambiamenti organizzativi avvenuti in un arco temporale piuttosto breve: in particolare alla community di sviluppatori è stata affiancata una fondazione con il compito di interfacciare i possibili nuovi utenti di MMbase. Inoltre, così come accade per buona parte dei prodotti OS "di successo", l'interesse del mondo business non ha tardato a manifestarsi: diversi *vendor*, hanno iniziato a distribuire il prodotto (creando delle distribuzioni personalizzate). Questa situazione è stata il fertile substrato sul quale sono germogliati numerosi conflitti tra gli attori (sviluppatori, fondazione, utenti, *vendor*) per diversi motivi portatori di interessi su MMbase. In particolare:

− gli sviluppatori sono un nucleo ristretto e molto coeso di tecnici, che sono entrati in conflitto sia con i *vendor* (accusati di scarsa collaborazione nello sviluppo condiviso della piattaforma) che con la stessa fondazione (percepita come un'entità che "si prende il merito" di MMbase, dato che è riuscita a costruire per il CMS un'immagine molto forte);
− i *vendor*: lamentano la scarsa collaborazione da parte degli sviluppatori, ad esempio nell'implementare in tempi accettabili funzionalità ritenute essenziali per il mercato;
− gli utenti: lamentano sia il *lock-in* da parte dei *vendor* (effettuato di fatto attraverso le distribuzioni), che la scarsa o insufficiente reattività da parte degli sviluppatori alle esigenze da loro evidenziate (gli utenti, come i *vendor*, sono aziende e quindi sono sottoposti a vincoli temporali e di mercato spesso molto stringenti, inoltre parte degli sviluppatori sono o sono stati pagati da loro)

Analogamente a quanto riferito da diversi studi, questi conflitti si stanno risolvendo grazie all'adozione di versioni di sviluppo "parallele". Nel caso di MMbase, ad esempio, convivono contemporaneamente una versione di sviluppo, una stabile (periodicamente ricreata da una di sviluppo) e le distribuzioni dei *vendor*, senza più alcune pretesa di uniformare i differenti approcci, ma con un continuo e proficuo scambio di informazioni tra i differenti attori.

## 8. Conclusioni

Quello del software è un mercato vero e proprio, difficilmente riducibile a "bene pubblico" senza rilevanti conseguenze dal punto di vista economico e occupazionale. È perciò necessario stabilire dei modelli di sostenibilità in grado di offrire adeguati ritorni economici anche a chi adotta modelli F/OSS, che si distinguono dai modelli di sviluppo di software tradizionale per la differente gestione della variabile "prezzo".
Il paradigma F/OSS, infatti, tende a sostituire la logica "*gift for reputation*" a quella "*good for money*", con la conseguenza di risultare spesso insufficientemente efficace nell'allocazione delle risorse. Questa inefficacia si esplicita in parecchi casi in un disallineamento tra domanda e offerta (i criteri e le priorità di sviluppo rischiano di non corrispondere a quelle che sono le reali esigenze e urgenze della potenziale domanda), eventualmente aggravata da fenomeni di virtuosismo, *happy engineerig* o mancanza di valore aggiunto per l'utilizzatore finale.

Da un punto di vista organizzativo si rileva che le motivazioni che spingono gli sviluppatori a partecipare a *community open source* e la tipologia di gratificazioni da essi ricercati sono legate alla crescita professionale e di reputazione. Tali aspetti dovrebbero essere tenuti in adeguata considerazione dalle *software house*, come strumenti atti a motivare i dipendenti e abbattere il *turnover*.
Alcuni autori, tuttavia, evidenziano come in alcuni casi il paradigma F/OSS possa fungere da freno all'innovatività.

Le comunità di sviluppatori, così come i ricercatori, hanno messo a punto alcuni modelli di business atti a garantire una certa sostenibilità economica a chi sviluppa F/OSS. È tuttavia importante ricordare che esistono tutta una serie di aree di attenzione (e tra le quali: necessità di disporre di sviluppatori permanenti, chiaro riconoscimento degli applicativi sviluppabili come *open source*, contestualizzazione del prezzo, ecc.) che le imprese che decidessero di basare il proprio business sul F/OSS dovrebbero tenere nella dovuta considerazione.

È altresì importante riconoscere il cambiamento paradigmatico che il F/OSS sta imponendo all'industria del software, al fine di sfruttare adeguatamente le opportunità che ne derivano. In particolare, l'adozione di un modello di sviluppo software basato su "*community* ibride" può consentire all'impresa lungimirante di ottenere vantaggi competitivi anche significativi (derivanti da: incremento di qualità del prodotto, diminuzione del TTM, contenimento dei costi, fidelizzazione del personale, ecc.). Tali vantaggi sono effettivamente ottenibili a patto che l'impresa sia in grado di comprendere a fondo l'impatto strutturale, culturale e sui processi derivante dalle peculiarità della filosofia F/OSS.

## Bibliografia